\documentclass[proof]{pasj00}
\draft

\begin{document}
\SetRunningHead{M. Shirahata et al.}{AKARI FIS Slow-Scan Calibration}
\Received{2008/10/08}%{yyyy/mm/dd}
\Accepted{2009/04/16}%{yyyy/mm/dd}

\title{Calibration and Performance of the AKARI Far-Infrared Surveyor (FIS) --- Slow-Scan Observation Mode for Point Sources}

%%% Begin:list of authors
% Do NOT capitalize all letters in "textsc".
%\author{A-Firtname \textsc{A-Familyname}},
%  \thanks{Example: Present Address is xxxxxxxxxx}}
%\affil{A-Address of Institute}
%\email{aaaaa@xxx.xxx.xx.xx}
%
%\author{B-Firstname \textsc{B-Familyname}}
%\affil{B-Address of Institute}\email{bbbbb@xxx.xxx.xx.xx}
%\and
%\author{C-Firstname {\sc C-Familyname}}
%\affil{C-Address of Institute}\email{ccccc@xxx.xxx.xx.xx}
%%% end:list of authors

%%% Please use the following style in case that sorting by 
%%% affilation is impossible. 
%
 \author{%
   Mai \textsc{Shirahata}\altaffilmark{1},
   Shuji \textsc{Matsuura}\altaffilmark{1},
   Sunao \textsc{Hasegawa}\altaffilmark{1},
   Takafumi \textsc{Ootsubo}\altaffilmark{1},
   Sin'itirou \textsc{Makiuti}\altaffilmark{1},
   Issei \textsc{Yamamura}\altaffilmark{1},
   Takao \textsc{Nakagawa}\altaffilmark{1},
   Mitsunobu \textsc{Kawada}\altaffilmark{2},
   Hidehiro \textsc{Kaneda}\altaffilmark{2},
   Hiroshi \textsc{Shibai}\altaffilmark{3,2},
   Yasuo \textsc{Doi}\altaffilmark{4},
   Toyoaki \textsc{Suzuki}\altaffilmark{5},
   Thomas \textsc{M$\rm{\ddot{u}}$ller}\altaffilmark{6},
   and
   Martin \textsc{Cohen}\altaffilmark{7}}
 \altaffiltext{1}{Institute of Space and Astronautical Science(ISAS), Japan Aerospace Exploration Agency(JAXA), 3-1-1 Yoshinodai, Sagamihara, Kanagawa 229-8510}
 \email{sirahata@ir.isas.jaxa.jp}
 \altaffiltext{2}{Graduate School of Science, Nagoya University, Furo-cho, Chikusa-ku, Nagoya 464-8602}
 \altaffiltext{3}{Graduate School of Science, Osaka University, 1-1 Machikaneyama-cho, Toyonaka-shi, Osaka 560-0043}
 \altaffiltext{4}{Department of General System Studies, Graduate School of Arts and Science, The University of Tokyo, 3-8-1 Komaba, Meguro-ku, Tokyo 153-8902}
 \altaffiltext{5}{National Astronomical Observatory of Japan, National Institutes of Natural Sciences, 2-21-1 Osawa, Mitaka, Tokyo 181-8588}
 \altaffiltext{6}{Max-Plank-Institut f$\ddot{u}$r extraterrestrische Physik, Giessenbachstrasse, 85748 Garching, Germany}
 \altaffiltext{7}{Radio Astronomy Laboratory, 601 Campbell Hall, University of California, Berkeley, CA 94720, USA}

%% `\KeyWords{}' always has to be placed before `\maketitle'.
 %Do NOT move this preamble from here!

\maketitle

\begin{abstract}
We present the characterization and calibration of the Slow-Scan observation mode 
of the Far-Infrared Surveyor (FIS) onboard the AKARI satellite.
The FIS, one of the two focal-plane instruments on AKARI, 
has four photometric bands between 50--180~$\rm{\mu m}$ 
with two types of Ge:Ga array detectors.
In addition to the All-Sky Survey, FIS has also taken 
detailed far-infrared images of selected targets by using the Slow-Scan mode.
The sensitivity of the Slow-Scan mode is one to two orders of magnitude 
better than that of the All-Sky Survey, 
because the exposure time on a targeted source is much longer. 
The point spread functions (PSFs) were obtained by observing 
several bright point-like objects such as asteroids, stars, and galaxies. 
The derived full widths at the half maximum (FWHMs) are 
$\sim$\timeform{30''} for the two shorter wavelength bands and 
$\sim$\timeform{40''} for the two longer wavelength bands, 
being consistent with those expected by the optical simulation, 
although a certain amount of excess is seen in the tails of the PSFs. 
The flux calibration has been performed
by the observations of well-established photometric calibration standards 
(asteroids and stars) in a wide range of fluxes. After establishing 
the method of aperture photometry, the photometric accuracy for point-sources 
is better than $\pm$15\% in all of the bands expect for the longest wavelength. 
\end{abstract}

\section{Introduction}\label{sec:intro}

AKARI is the first Japanese satellite dedicated to infrared astronomy (Murakami et al. 2007). 
It was successfully launched on February 21st in 2006 (UT) from the Uchinoura Space Center (USC) 
with the M-V rocket of Japan Aerospace Exploration Agency (JAXA). 
AKARI opened its aperture lid to begin astronomical observation on April 13th 2006. 
AKARI achieved continuous observation during 550~days 
by the time when the liquid-helium ran out on August 26th 2007. 
AKARI was thrown in a sun-synchronous polar orbit 
along the twilight zone at an altitude of $\sim$700~km 
in order to carry out the All-Sky Survey in every half-year. 
The telescope has the size of 68.5~cm in diameter (Kaneda et al. 2007), 
and is cooled down at lower than 6~K in a liquid-helium cryostat (Nakagawa et al. 2007)
for reduction of thermal emission from the instruments. 

The Far-Infrared Surveyor (FIS) is one of the two focal-plane instruments onboard AKARI (Kawada et al. 2007). 
It has four photometric bands between 50--180~$\rm{\mu m}$, with two types of Ge:Ga array devices; 
the Short-Wavelength (SW) detector and the Long-Wavelength (LW) detector. 
The SW detector (Fujiwara et al. 2003), 
which is responsible for the wavelength range in 50--110~$\rm{\mu m}$, 
is a two-dimensional monolithic Ge:Ga photoconductor array. 
This array is directly connected to a cryogenic readout electronics (CRE) by Indium-bumping technology. 
This detector was developed under collaboration among 
the National Institute of Information and Communications Technology (NICT), 
Institute of Space and Astronautical Science (ISAS) of Japan Aerospace Exploration Agency (JAXA), 
and Nagoya University. 
On the other hand, the LW detector (Doi et al. 2002) for 110--180~$\rm{\mu m}$ 
is a compact stressed Ge:Ga photoconductor array having a 
special cavity structure and a stressing mechanism. 
Ge:Ga chips for the LW detector were supplied by NICT, 
and the LW detector module was assembled by the University of Tokyo. 
As for the CREs of the SW and LW detectors, 
we developed a Capacitive Trans-Impedance Amplifier (CTIA) with 
silicon p-MOSFETs optimized for low-temperature operation (Nagata et al. 2004). 
FIS was designed primarily to perform the All-Sky Survey, 
which is the main purpose of the AKARI mission. 
Additionally, FIS has two other operation modes with pointing attitude control; 
the Slow-Scan mode to take detailed images of specific targets 
and the spectroscopy mode to take spectra by using the Fourier transform spectrometer 
(Kawada et al. 2008; Murakami et al. 2009, in prep.). 
The sensitivity in the Slow-Scan mode is 1--2 orders of magnitude 
better than that of the All-Sky Survey, 
because the exposure time on a target source is longer than the All-Sky Survey mode. 
In this paper, we describe the data reduction procedures and 
the in-orbit performance of the Slow-Scan mode, 
the imaging performance, and the flux calibration. 

Characterization and calibration of FIS were performed by observing 
two kinds of well-established photometric calibration standards. 
One of them are solar-system objects such as asteroids and planets, 
and the others are stars. 
These calibration sources were studied well enough 
(asteroids; M$\rm{\ddot{u}}$ller \& Lagerros 1998, 2002, stars; Cohen et al. 1999, 2003a, 2003b), 
and were used as the calibration standard in the previous infrared missions 
(e.g., ISO/ISOCAM; Blommaert et al. 2003, ISO/ISOPHOT; Schulz et al. 2002, 
COBE/DIRBE; Hauser et al. 1998, and Spitzer/MIPS; Gordon et al. 2007, Stansberry et al. 2007). 
Some of the galaxies observed by ISO and IRAS were also examined 
not only for the cross calibration but also for the calibration check 
for the objects with much redder spectra than stars and asteroids. 

In Sec.~2, we explain the method of the FIS Slow-Scan observation. 
We describe observations of the calibration standards in Sec.~3. 
Careful data reduction procedures are required in order to obtain high-quality images 
and to establish the accurate calibration. 
Accordingly, we describe the process of data reduction in Sec.~4. 
The imaging performance is discussed in Sec.~5. 
In Sec.~6, we describe the calibration factors, 
and their flux dependency by considering the characteristics of the Ge:Ga detectors. 
We also checked the consistency of the calibration for various observations. 
Finally, we summarize our results in Sec.~7. 
The calibration factors derived by the present work are 
the official values of AKARI/FIS Slow-Scan calibration.

\section{FIS Slow-Scan Observation}

\begin{figure}
  \begin{center}
    \FigureFile(80mm,80mm){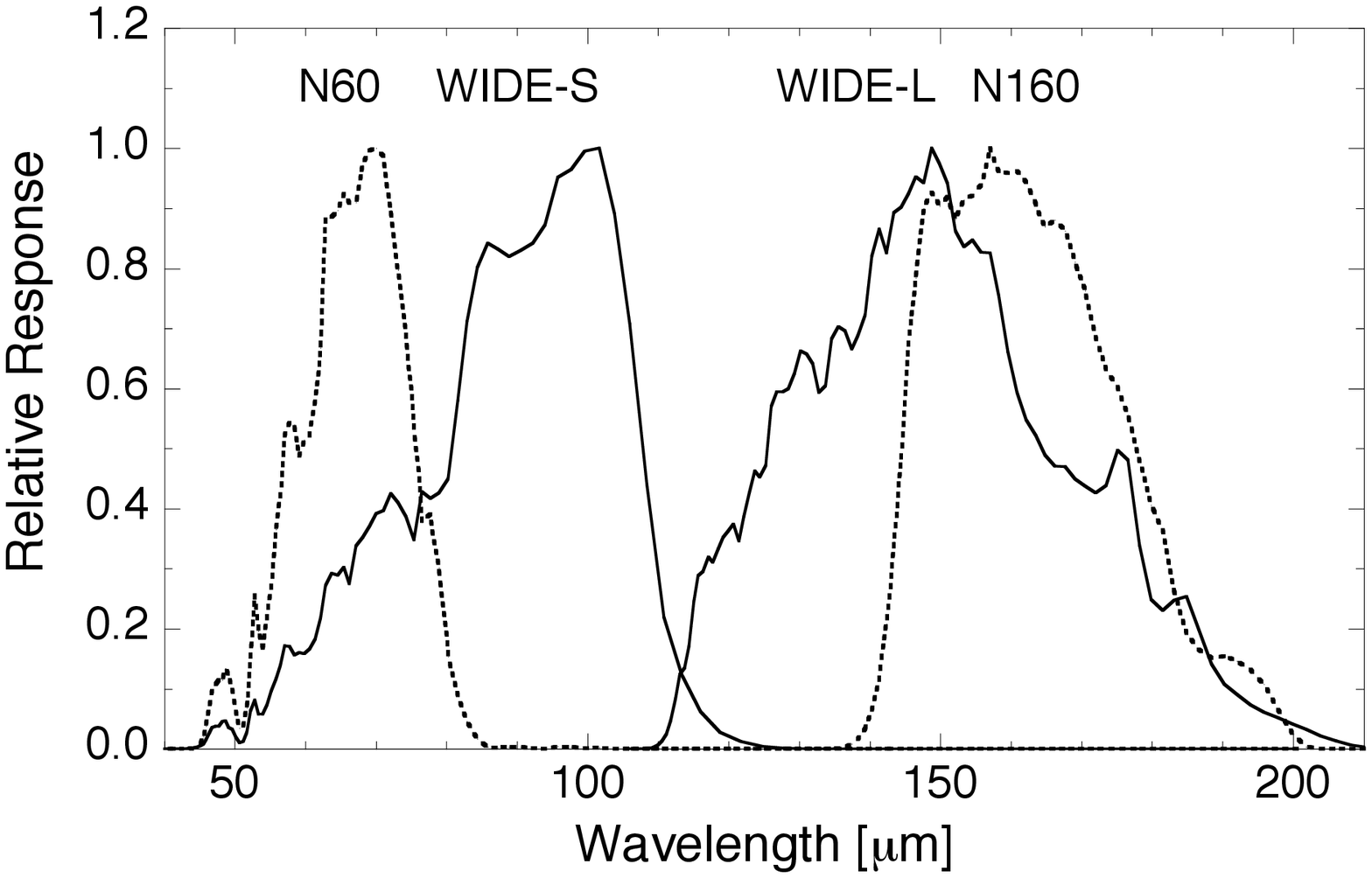}
  \end{center}
  \caption{
    The system spectral response of the FIS photometric bands. 
    These profiles were derived from the spectral response of 
    optical components and the detector spectral responsivity. 
    These profiles provide typical band shape, 
    because detectors have the pixel-to-pixel variation of the spectral responsivity. 
    }\label{fig:rsrf}
\end{figure}

The FIS has four photometric bands 
whose center wavelengths are represented at 65, 90, 140, and 160~$\rm{\mu m}$. 
These band shapes, shown in Fig.~\ref{fig:rsrf}, 
are formed by the combination of optical filters and two detector devices. 
Each detector device covers one broad-band and one narrow-band; 
{\it WIDE-S} (90~$\rm{\mu m}$) and {\it N60} (65~$\rm{\mu m}$) for the SW detector 
and {\it WIDE-L} (140~$\rm{\mu m}$) and {\it N160} (160~$\rm{\mu m}$) for the LW detector. 
The array formats are 3$\times$20, 2$\times$20, 3$\times$15, and 2$\times$15 
for the {\it WIDE-S}, {\it N60}, {\it WIDE-L}, and {\it N160} bands, respectively. 
The pixel scales are 26.8~arcsec for the SW detector and 44.2~arcsec for the LW detector, 
whose values are comparable to the diffraction limits of the telescope. 
The field-of-views (FOVs) of two detector devices overlap on the sky. 
The specifications for the FIS instrument are summarized in Tab.~\ref{tab:fis}. 

As reported by many authors, 
Ge:Ga detectors show slow transient response at low temperature 
under low background flux conditions, i.e., in the space environment 
(Kaneda et al. 2002, Hiromoto \& Fujiwara 1999, Haegel et al. 1999, Fouks \& Schubert 1995, and references therein). 
This slow transient response indicates time-delay in response for 
quick changes of the incoming flux. 
The time constant of the slow response is typically 10--100~seconds, 
and it depends on both the background and the signal photon fluxes. 
Therefore, in the case of AKARI which conduct the scan observations, 
the slow response causes decrease of the sensitivity for compact sources. 
In order to reduce the effects of the slow response of the LW detector, 
an offset light source is placed in front of the LW detector unit 
to irradiate each detector pixels with a constant intensity. 
The intensity of this offset light source was set to be $\sim$100~$\rm{MJy~sr^{-1}}$ 
to obtain the highest signal-to-noise ratio 
taking the increase of the photon noise into account. 

The FIS Slow-Scan observations are carried out using 
the AKARI astronomical observation templates (AOTs) FIS01 and FIS02. 
FIS01 is designed for compact source photometry with two sets of round-trip scans, 
while FIS02 is suitable for wide area mapping with a long single round-trip scan. 
Observers can select the scan speed (\timeform{8''}~s$^{-1}$ or \timeform{15''}~s$^{-1}$) 
and the reset interval (0.5~sec, 1.0~sec, or 2.0~sec). 
In case of FIS01, the cross-scan shift length between two round trips 
is also selectable from either a few pixels (\timeform{70''}) or half of the FOV (\timeform{240''}). 
More details about the AOTs are described in Kawada et al. (2007) and Verdugo et al. (2007). 

One set of the Slow-Scan observation takes about 30~minutes, 
including the calibration sequence. 
The dark current measurement and responsivity check with an internal calibrator are executed 
by closing the cold shutter during 7~minutes of the attitude maneuver operation 
between the All-Sky Survey position and the target position of the pointing observation. 
After the maneuver completes, the shutter is opened for monitoring 
the sky signal during the settling time for 
fine control of the satellite attitude, and then the Slow-Scan starts. 
At the turning point(s) of the round-trip scan(s), the shutter is briefly closed 
and the internal calibration lamps are turned on to monitor the time variation of the detector responsivity. 
The total observation time for the round-trip scan(s) is about 12~minutes. 
The same calibration sequence is repeated 
during the maneuver for returning to the All-Sky Survey mode after the pointing observation. 
Such a highly-redundant calibration data set enabled us to correct 
the response changes of the detectors, 
referring to astronomical calibration data taken by separate observations with the same calibration sequence.

\section{Targets}

For evaluation of the imaging performance and flux calibration, 
several kinds of astronomical point-like sources were observed; 
18~Slow-Scan observations of 14~stars, 
17~Slow-Scan observations of 11~asteroids, 
2~Slow-Scan observations of a planet (Neptune), 
and 13~Slow-Scan observations of 11~galaxies. 
All observations were carried out by the FIS01 (two round-trip scans) mode 
with a \timeform{70''} shift. 
The scan speed was \timeform{8''}~$\rm{s^{-1}}$ for most cases to obtain a high signal-to-noise ratio. 
A few observations were performed with the \timeform{15''}~$\rm{s^{-1}}$ scan 
in order to check the influence caused by the differences of the scan speed. 
The reset interval was chosen so as to avoid saturation for each target. 

The calibration stars were selected from the standard star catalog 
for spectral types of $AV$ and $KIII$ established by M. Cohen et al. (1999, 2003a, 2003b). 
These stars have been studied extensively and are reliable as the calibration standards 
with the accuracies better than 6\%. 
Unfortunately, the stars are too faint to use as calibrators 
for the {\it WIDE-L} and {\it N160} bands. 

Since asteroids are brighter than stars, they are 
widely used as the far-infrared calibrators (M$\rm{\ddot{u}}$ller \& Lagerros 2003). 
We used them for the {\it N60} and {\it WIDE-S} bands and also 
for the {\it WIDE-L} and {\it N160} bands. 
Asteroids show flux variations on various levels and time scales 
due to the change of illumination, observing geometries, and also their rotation. 
Therefore, we estimated the expected flux for each observation 
from the thermophysical model (TPM) (M$\rm{\ddot{u}}$ller \& Lagerros 1998, 2002) 
by taking the geometries and surface properties into account. 
The flux uncertainties (see Tab.~\ref{tab:obslog}) are based on a thorough 
analysis of the TPM input parameters and the number and 
quality of well-calibrated thermal observations for each asteroid. 
The apparent motions of the asteroids are typically less than \timeform{1'}~$\rm{hour^{-1}}$, 
and are taken into account at image co-addition (see Fig.~\ref{fig:image}). 

We also observed several point-like galaxies (luminous infrared galaxies) 
whose fluxes have been well studied by IRAS and ISO. 
We used bright samples for the measurement of 
the point spread functions (PSFs) and the encircled energy functions, 
especially for the {\it N160} band. 

Table~\ref{tab:obslog} summarizes the observation log and the expected fluxes with their uncertainties. 
In all the four bands, the expected fluxes range from 
$\sim$100~mJy to $\sim$100~Jy, i.e., more than three orders of magnitudes. 
Note that the expected fluxes listed here are color-corrected values by assuming 
the flat spectrum (i.e., $\nu F_{\nu}=$~const.). 
All observations had been carried out in the performance verification (PV) phase 
and in the engineering time available for calibration which is occasionally allocated 
according to the requirement.

\section{Data Reduction}

The raw data were processed using the official FIS slow-scan data-analysis toolkit (SS-tool) 
to produce the time-series calibrated data and the final co-added map. 
More details about the standard reduction steps are described in Matsuura et al. (2007). 

First of all, the raw data were processed by the software tools to do 
ADU-to-volt conversion, flagging of bad data (dead pixels, saturated pixels, reset anomalies, 
and other discontinuities) and the correction of non-linear integration ramps. 

The next process is slope calculation of each ramp, 
removing the cosmic-ray events identified in the ramp. 
This process maximizes the signal-to-noise ratio and avoids any periodic structure 
arising from incompleteness of the non-linear ramp correction. 
The reduced data provides 1--10~samples per pixel for a source crossing time, 
which depend on both the scan speed and the reset interval. 
By co-adding the data of all array pixels, 
the Nyquist sampling condition is satisfied in the real-space domain. 
Glitches and subsequent tails induced by cosmic-ray hits affect 
the data severely (Suzuki et al. 2008). 
The integration ramps affected by the glitches were eliminated in the slope calculation process. 
The tails were not flagged-out at this stage, but 
the affected data were removed by sigma clipping in the co-addition process. 

The third step of the SS-tool processing is to produce calibrated data 
for each array pixel after dark current subtraction and flat-fielding. 
The pixel-to-pixel variations of the detector responsivity are approximately 
15\% and 50\% of the average for the SW detector and the LW detector, respectively. 
Therefore, the correction of the responsivity variation is necessary to obtain accurate images. 
This process is based on the FIS observations of known diffuse sources: 
zodiacal emissions and interstellar dust emissions, which are expected 
to be smooth within the field of view. 
In this process we can also compare the sky brightness measured by the FIS with DIRBE/COBE, 
which provide a well calibrated far-infrared sky map (Matsuura et al., in prep.). 
Accordingly, the unit conversion from the instrumental unit to the surface brightness 
is simultaneously performed. 

The zodiacal emission is expected to be almost perfectly flat; 
their anisotropies on arcminute scales are less than 1\% (Abraham et al. 1997). 
In the cases of the {\it N60} and {\it WIDE-S} bands, 
the sky brightness of almost entire sky except for the Galactic Plane regions 
is dominated by the zodiacal emission, and the contributions of 
galactic cirrus (interstellar dust) emission are expected to be negligible. 
Hence, the blank sky observed near the target source can be used for flat-fielding, 
as long as there is no bright source in there. 
The response distribution of the detector array pixels is derived from 
the average of time-series data during the Slow-Scan, where the average is 
calculated after removing data that exceed the 3-$\sigma$ noise level for each pixel. 
The flat is built based on sky monitoring data acquired 
during the attitude-settling time just before the Slow-Scan observation. 
The flat-fielding is done by dividing the data by the response distribution. 
With this 'self' flat-fielding method, any stripes in the image caused by the flat field errors 
are buried under the random noise. 

In the cases of the {\it WIDE-L} and {\it N160} bands, 
the detector signals are dominated by offset light 
implemented to improve the slow transient response of the stressed Ge:Ga. 
Although galactic cirrus emissions at high latitudes could be a flat source 
with moderate smoothness ($\sim$10\%) and relatively high brightness, 
their signals are less than 10\% of that of the offset light. 
The intensity distributions of the offset light at the aperture of the detector arrays 
are estimated from laboratory measurements to be uniform to within 10\%. 
Hence, the 'self' flat-fielding method with an average sky signal including offset light 
was applied to correct the responsivity variation in the detector arrays, 
as was done for the {\it N60} and {\it WIDE-S} bands. 

The final step of the SS-tool processing is co-addition of the calibrated time-series data 
onto a sky grid. The sky position of each data point is derived from the telescope boresight 
according to the satellite attitude and from the array pixel map on the focal-plane. 
The grid pixel sizes were set to \timeform{2.5''}. 
A sufficient number of data point per grid pixel ($>$5 on average) is secured 
by considering the finite size of the detector array pixel (Drizzle method; Williams et al. 1996). 
In the co-adding process, small glitches and other artifacts were sigma-clipped 
with the standard deviation calculated at each grid pixel. 
The threshold for the sigma-clipping was set to 5-times the standard deviation. 
The fraction of the rejected data points in this process was less than 1\% of the original data. 

Figure~\ref{fig:image} shows an example of the final co-added images 
(asteroid Ceres) for all the FIS bands. 
The imaging performance and flux calibration based on the obtained maps 
are discussed in the following section.

\section{Imaging Performance}

\subsection{Point Spread Functions (PSFs)}

\begin{figure}
  \begin{center}
    \FigureFile(80mm,80mm){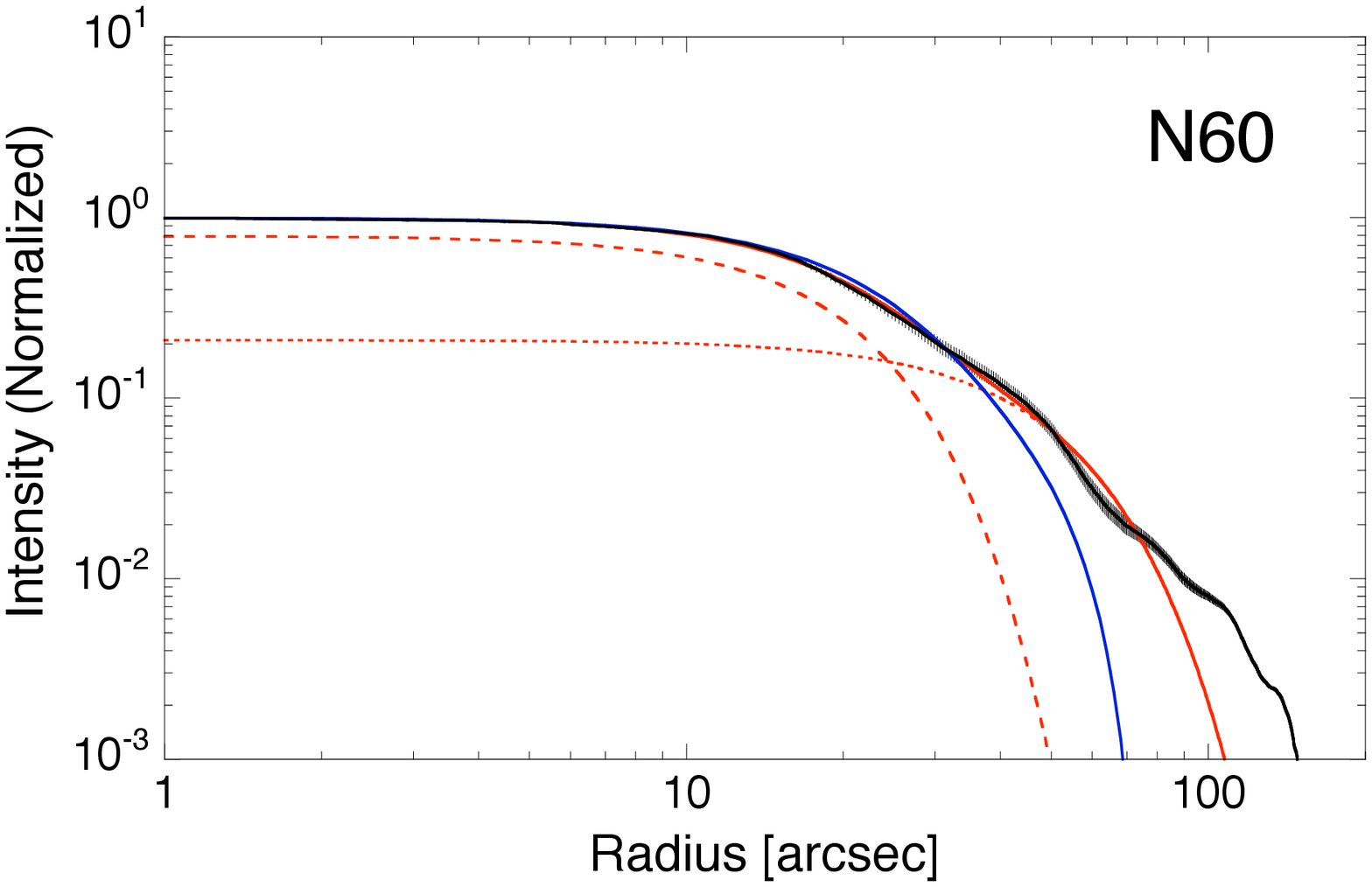} \quad
    \FigureFile(80mm,80mm){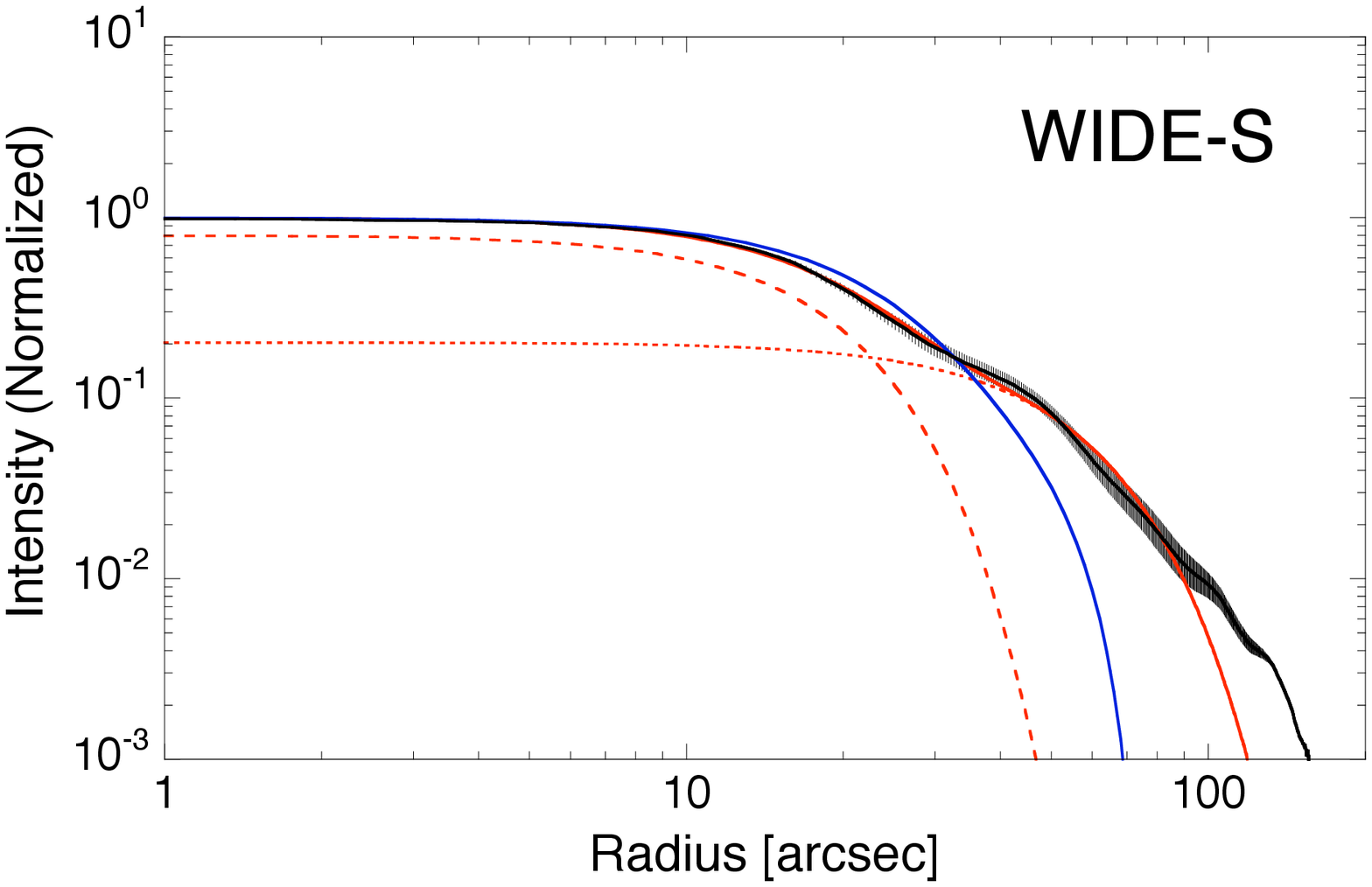}
    \FigureFile(80mm,80mm){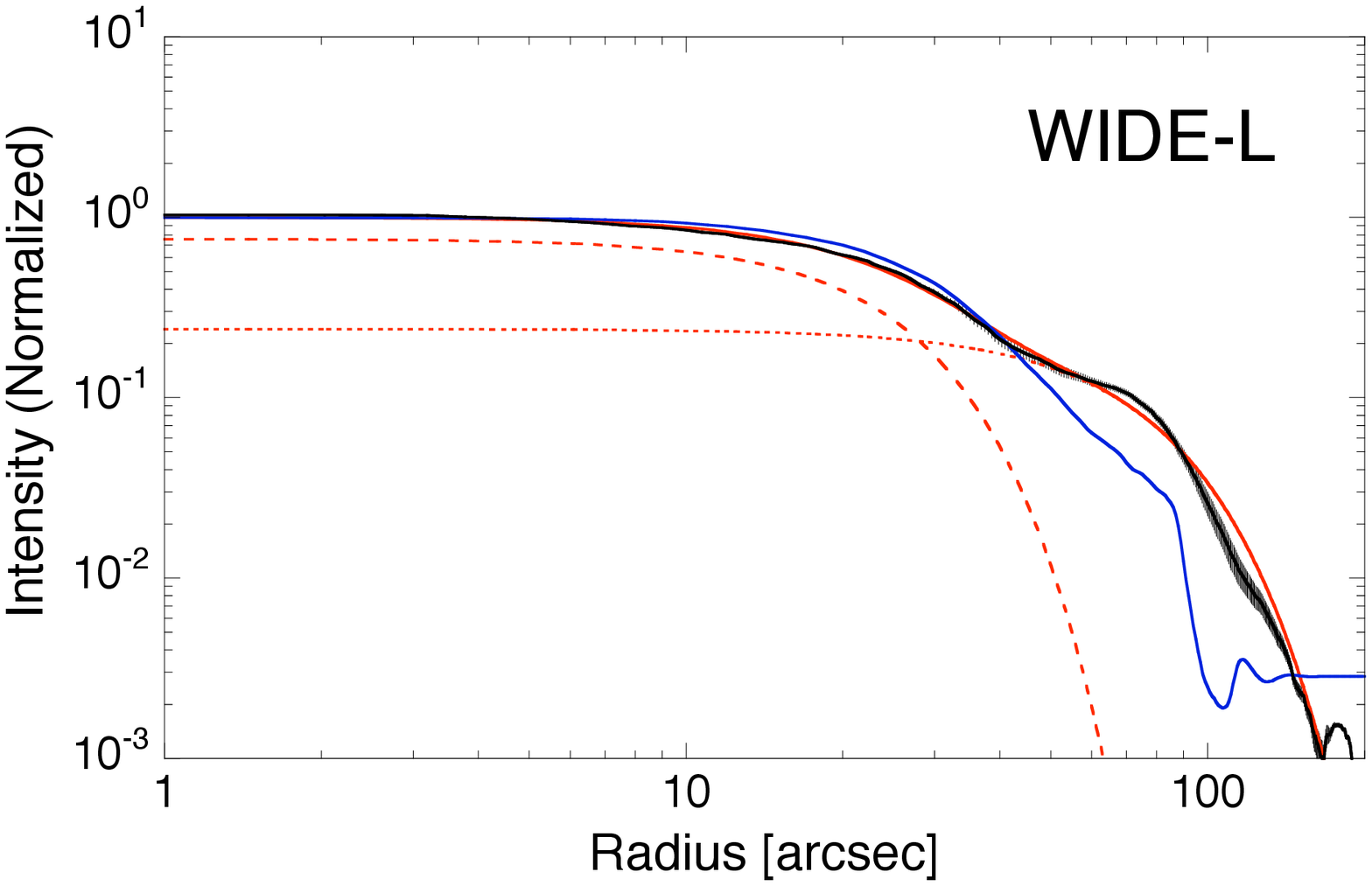} \quad
    \FigureFile(80mm,80mm){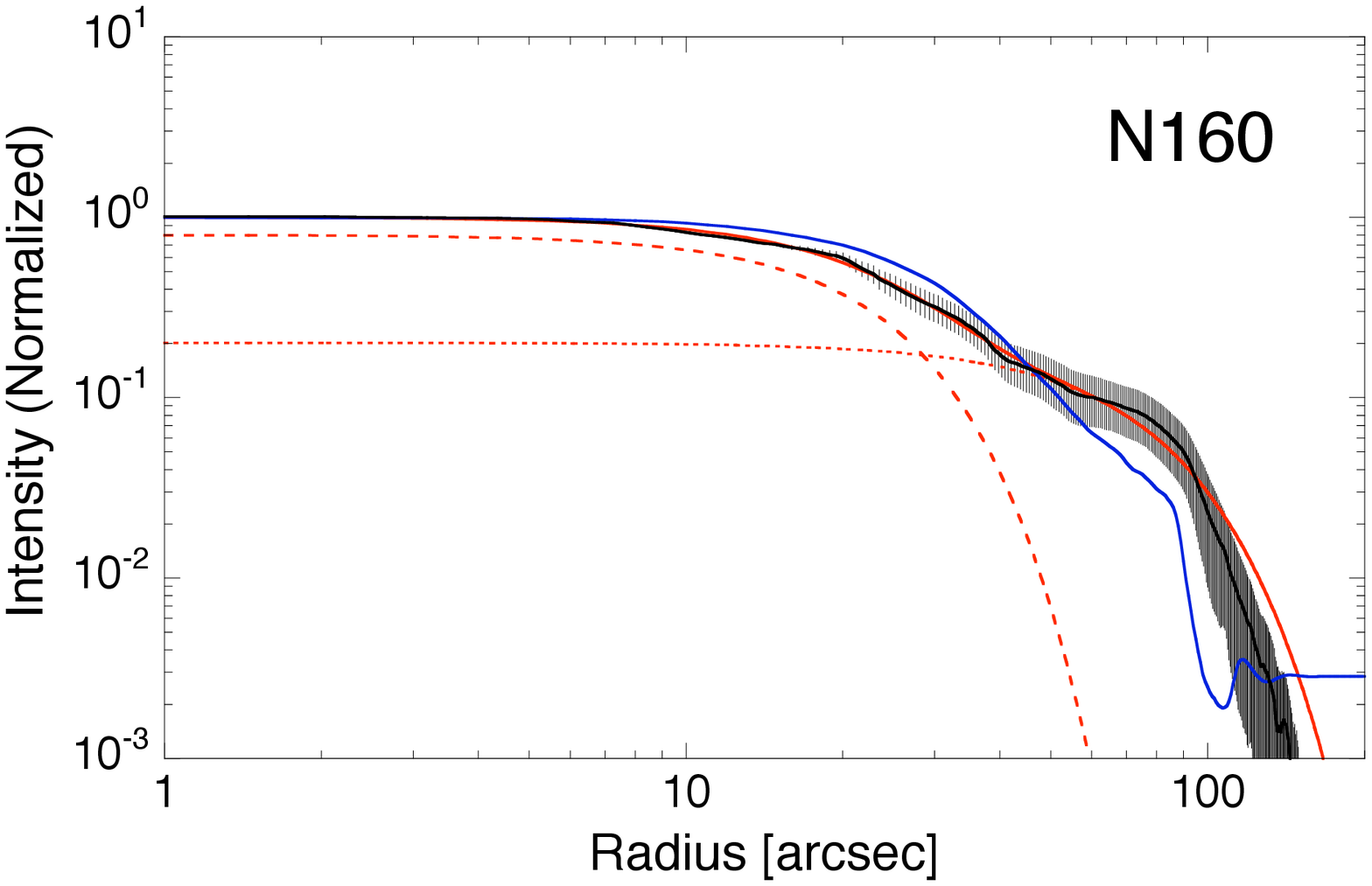}
  \end{center}
  \caption{
    Radial profiles of the observed PSFs and the optical simulation model. 
    The observed profiles (black solid lines) are derived from the observation of bright point-sources. 
    The error bars indicate the standard deviation. 
    The results of the two component Gaussian fitting are represented 
    by red solid lines (total) and red dotted lines (each component). 
    The expected profiles from the optical simulation model are shown by blue solid lines. 
    }\label{fig:psf}
\end{figure}

Before the launch of AKARI, we checked the PSFs 
in the laboratory using a pin-hole source located on 
the focal-plane of the FIS optics (Shirahata et al. 2003). 
The measured PSFs were almost consistent with those expected from the optical simulation, 
though the uncertainty originated in the measurement system was somewhat large. 
The PSFs, throughout the optics including the AKARI telescope, 
were evaluated in orbit by observations of bright point-sources. 
These sources are chosen with the condition that 
S/N for single pointing observation is better than 300, 
which are designated in Tab.~\ref{tab:obslog} by the bold-faced type or the circle symbols; 
19 sources brighter than 10~Jy for the {\it N60} band, 
28 sources brighter than 2~Jy for the {\it WIDE-S} band, 
6 sources brighter than 20~Jy for the {\it WIDE-L} band, 
and 4 sources brighter than 50~Jy for the {\it N160} band. 
Figure~\ref{fig:psf} shows the azimuthally-averaged radial profiles 
of the measured PSFs and compares them to the optical simulation model (Jeong et al. 2003).
The averaged profiles are well reproduced by the two-component Gaussian functions, 
whose parameters are summarized in Tab.~\ref{tab:psf}. 
The full widths at the half maximum (FWHM) of the main Gaussian component are 
$32\pm1$, $30\pm1$, $41\pm1$, and $38\pm1$~arcsec 
for the {\it N60}, {\it WIDE-S}, {\it WIDE-L}, and {\it N160} bands, respectively. 
At the tails of the PSFs, there are significant excess 
compared with the optical simulation, 
whose contributions are about 20\% of the total power. 
We did not see any significant dependence in the PSFs on the infrared color of the targets.

\subsection{Cross-talk and Ghost Signals}

\begin{figure}
  \begin{center}
    \FigureFile(160mm,160mm){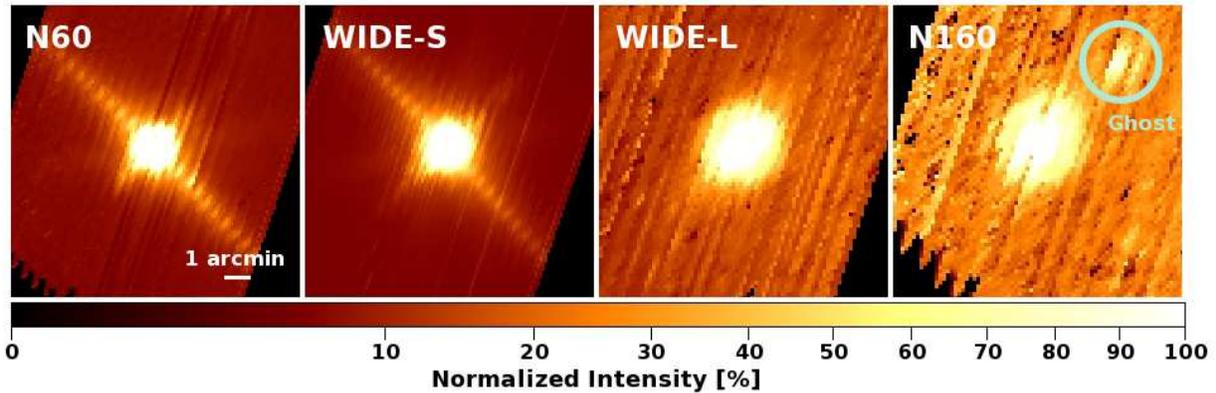}
  \end{center}
  \caption{
    Examples of the final co-added image obtained by the FIS Slow-Scan observation mode (AOT FIS01). 
    The panels show {\it N60}, {\it WIDE-S}, {\it WIDE-L}, and {\it N160} images from left to right. 
    The image size is \timeform{10'} $\times$ \timeform{10'}. 
    The color scale is square root contours. 
    In the images of the {\it N60} and {\it WIDE-S} bands, cross-talk signals are seen 
    along both array axes. 
    Ghost signals are seen in all bands, especially in the {\it N160} band. 
  }\label{fig:image}
\end{figure}

One property affecting the imaging quality is cross-talk between array pixels. 
This effect is seen only with the SW detector, which has a monolithic structure. 
In the maps of the SW detector ({\it N60} and {\it WIDE-S}) shown in Fig.~\ref{fig:image}, 
cross-talk signals are clearly seen along the detector array axes. 
Amplitude of the cross-talk signal is roughly 5\% or less of the peak signal. 
Possibility to explain this phenomenon is carrier diffusion or 
internal-reflection in the Ge:Ga substrate. 

Another concern in terms of imaging quality is the presence of ghost signals. 
A ghost image appears in the all bands near the target image, 
and is prominent especially for the {\it N160} band as shown in Fig.~\ref{fig:image}. 
The center positions of broad-band and narrow-band arrays have an angular separation 
of $\sim$6~arcmin in the scanning direction, and the ghost signal of each array 
appears separate time corresponding to the angular separation. 
Therefore, the ghost signal appears in one array when a strong light 
enters the other array of the same detector. 
The cause of the ghost is presumably 
electrical cross-talk in the multiplexer of the cryogenic readout electronics. 
The position and intensity of the ghost signal to the target signal, 
which are summarized in Tab.~\ref{tab:ghost}, 
had been stable throughout the mission. 
As long as the target is a point-source, the ghost should be removable.

\section{Flux Calibration}

\subsection{Calibration Factor}

As described in Sec.~4, the final co-added maps produced by the SS-tool 
have units of surface brightness in MJy~$\rm{sr^{-1}}$, 
which has been calibrated in the ealier stage in the data processing 
by diffuse sources such as the zodiacal emission and the interstellar dust emission (Matsuura et al. in prep.). 
The application of aperture photometry for the co-added map gives 
the temporal flux of the target source. 
Hereafter, we derive the calibration factors for point-source photometry 
based on the observations of the calibration standards. 

For the source extraction and the aperture photometry, 
we used the photometry tools FIND, GCENTRD, and APER 
in the IDL Astronomy User's Library (Landsman 1993) at NASA/GSFC. 
These tools search for the center positions of point-sources 
with a Gaussian window function of the same width as the PSF, 
and measure the fluxes by integrating the pixel values within the aperture. 
The background sky levels were estimated to be the average of the surface brightness in the sky annulus, 
defined as \timeform{2.3'}--\timeform{3.3'} in radius for the {\it WIDE-S} and {\it N60} bands, 
and \timeform{3.0'}--\timeform{4.0'} for the {\it WIDE-L} and {\it N160} bands, respectively. 
The inner radius of the sky annulus is distant enough from the target source, and 
the outer radius is distant enough from the edges of the maps. 

\begin{figure}
  \begin{center}
    \FigureFile(80mm,80mm){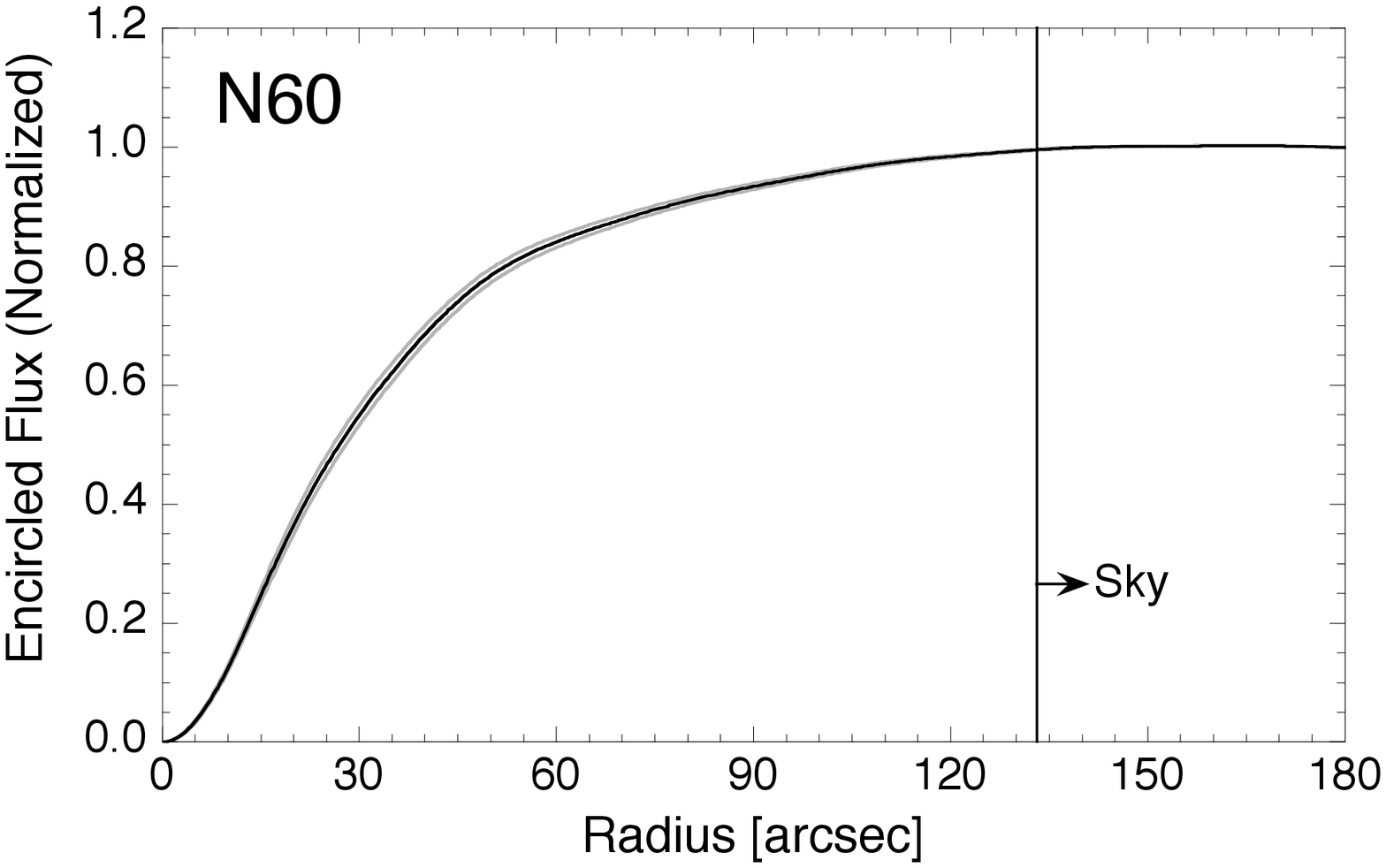} \quad
    \FigureFile(80mm,80mm){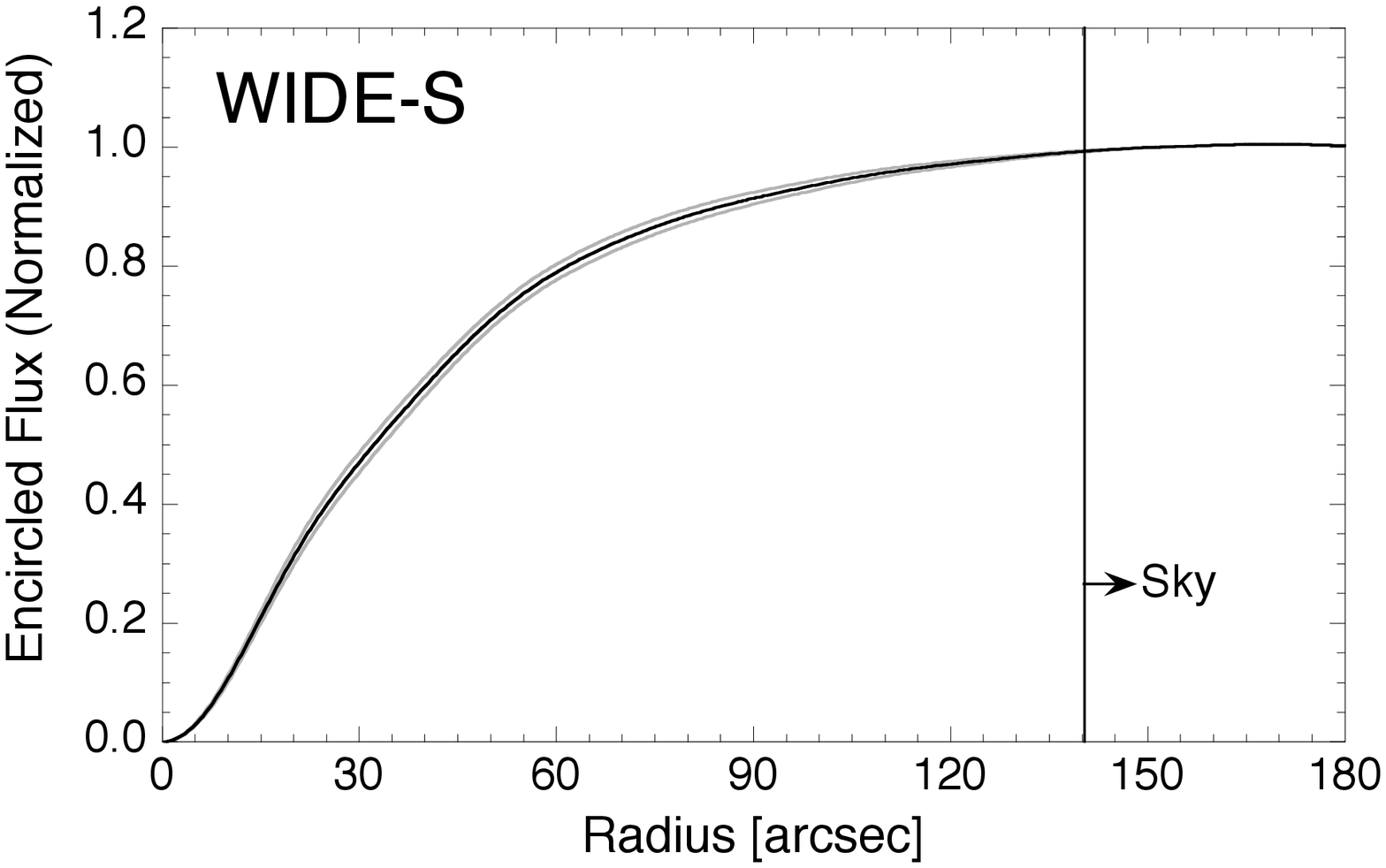}
    \FigureFile(80mm,80mm){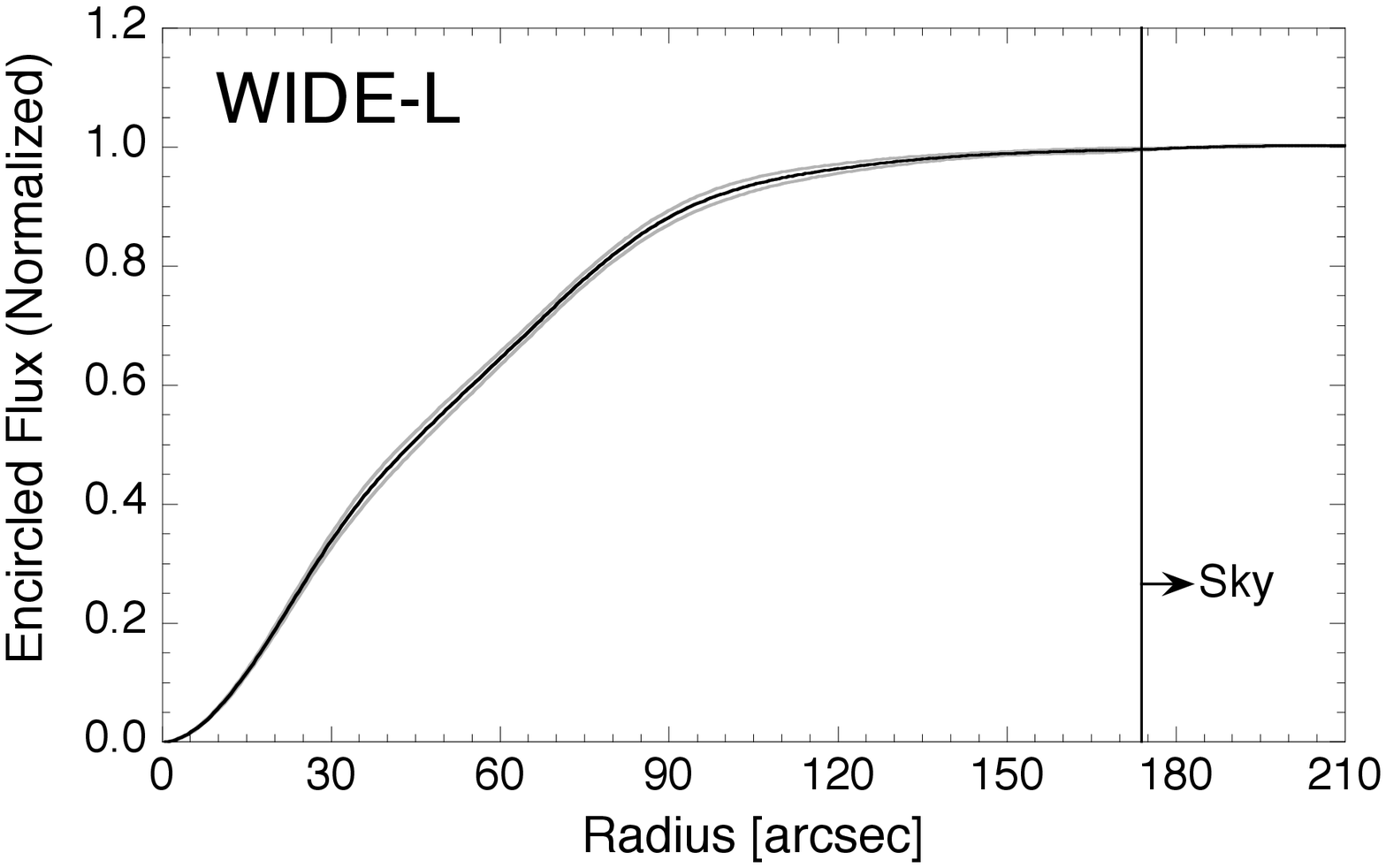} \quad
    \FigureFile(80mm,80mm){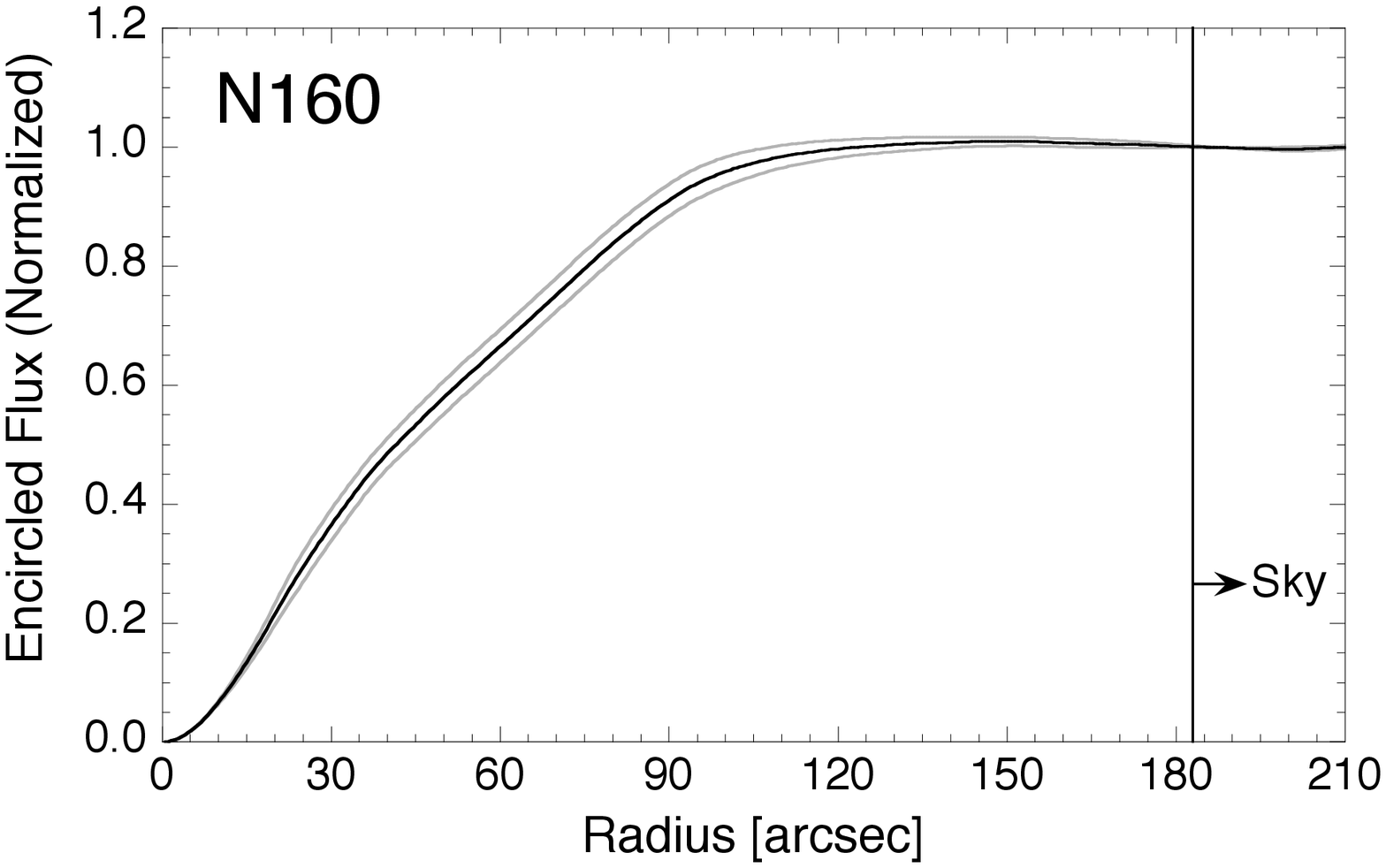}
  \end{center}
  \caption{
    Encircled energy function. 
    The gray region indicates the range of the standard deviation of sample. 
    All profiles are normalized to 1.0 at the sky region.
    }\label{fig:cog}
\end{figure}

Before performing the aperture photometry for all of the calibration standards, 
we compounded the encircled energy functions from the images of 
the bright sources that were used to measure the PSFs. 
First we computed the encircled energy for each source with various aperture radii, 
and then normalized the result at the background sky flux level. 
The obtained profiles of the encircled energy show good agreement with each other, 
within $\sim$3\% for the shorter 3-bands 
and $\sim$6\% for the {\it N160} band, as shown in Fig.~\ref{fig:cog}. 
The aperture correction factors as the normalized encircled energy functions 
are summarized in Tab.~\ref{tab:cog}. 

We performed aperture photometry on all images of the calibration standards 
by using the aperture radius of 
\timeform{40''} for the {\it WIDE-S} and {\it N60} bands, and 
\timeform{60''} for the {\it WIDE-L} and {\it N160} bands, respectively, 
and applied the aperture correction with the factors tabulated in Tab.~\ref{tab:cog}. 
These aperture radii are large enough to minimize uncertainties due to 
centroiding errors and to ensure that any uncertainties in the 
aperture correction have a small effect on the derived fluxes. 
In addition, these aperture sizes enable us to improve the S/N of the photometry 
and thereby to extend the calibration to the sources with somewhat fainter flux densities. 
We derive the uncertainty on each measurement from the scatter of the pixel values 
in the background annulus, which contributes both the uncertainty due to 
summing the object flux density as well as subtracting the background. 

\begin{figure}
  \begin{center}
    \FigureFile(80mm,80mm){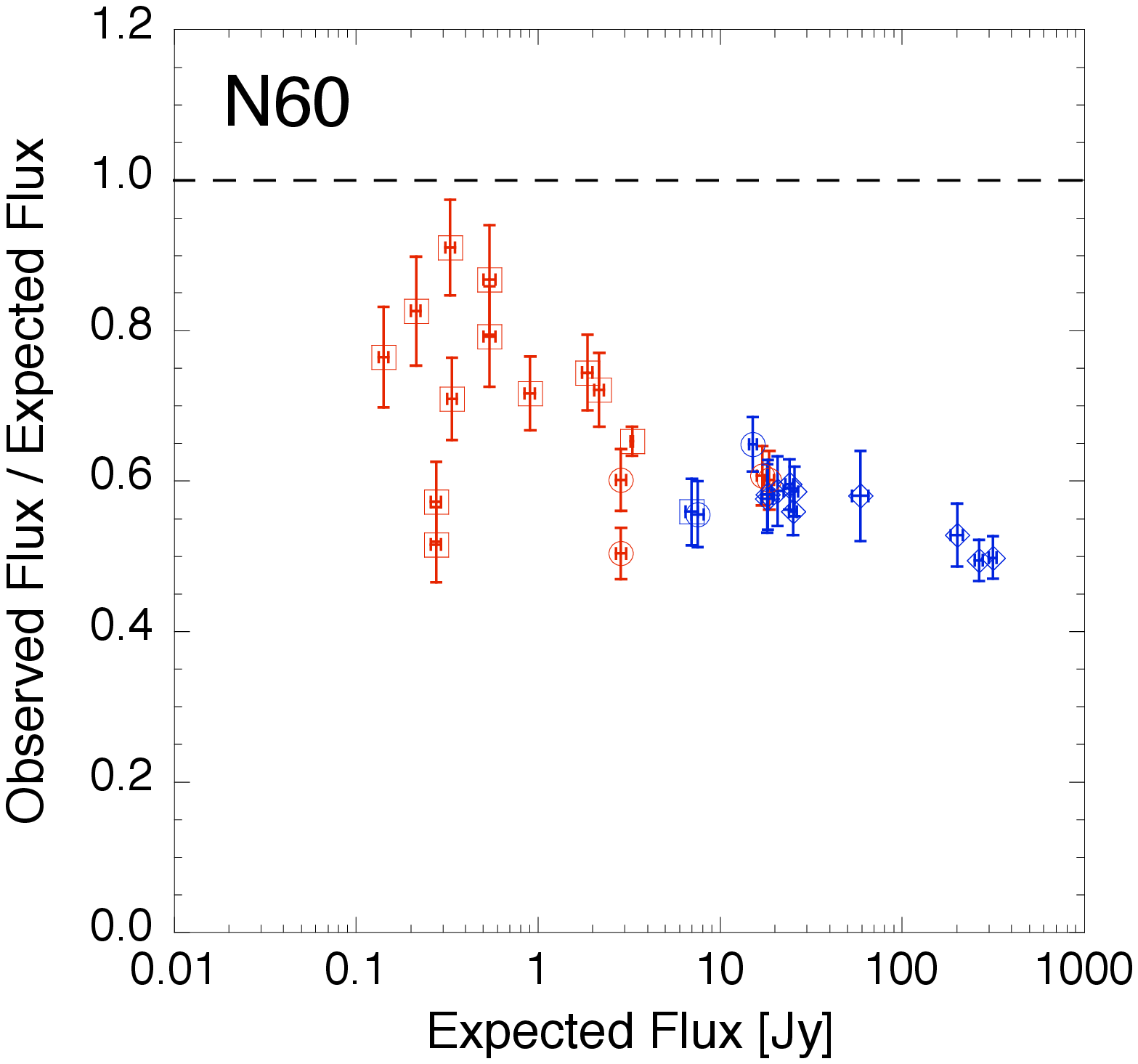} \quad
    \FigureFile(80mm,80mm){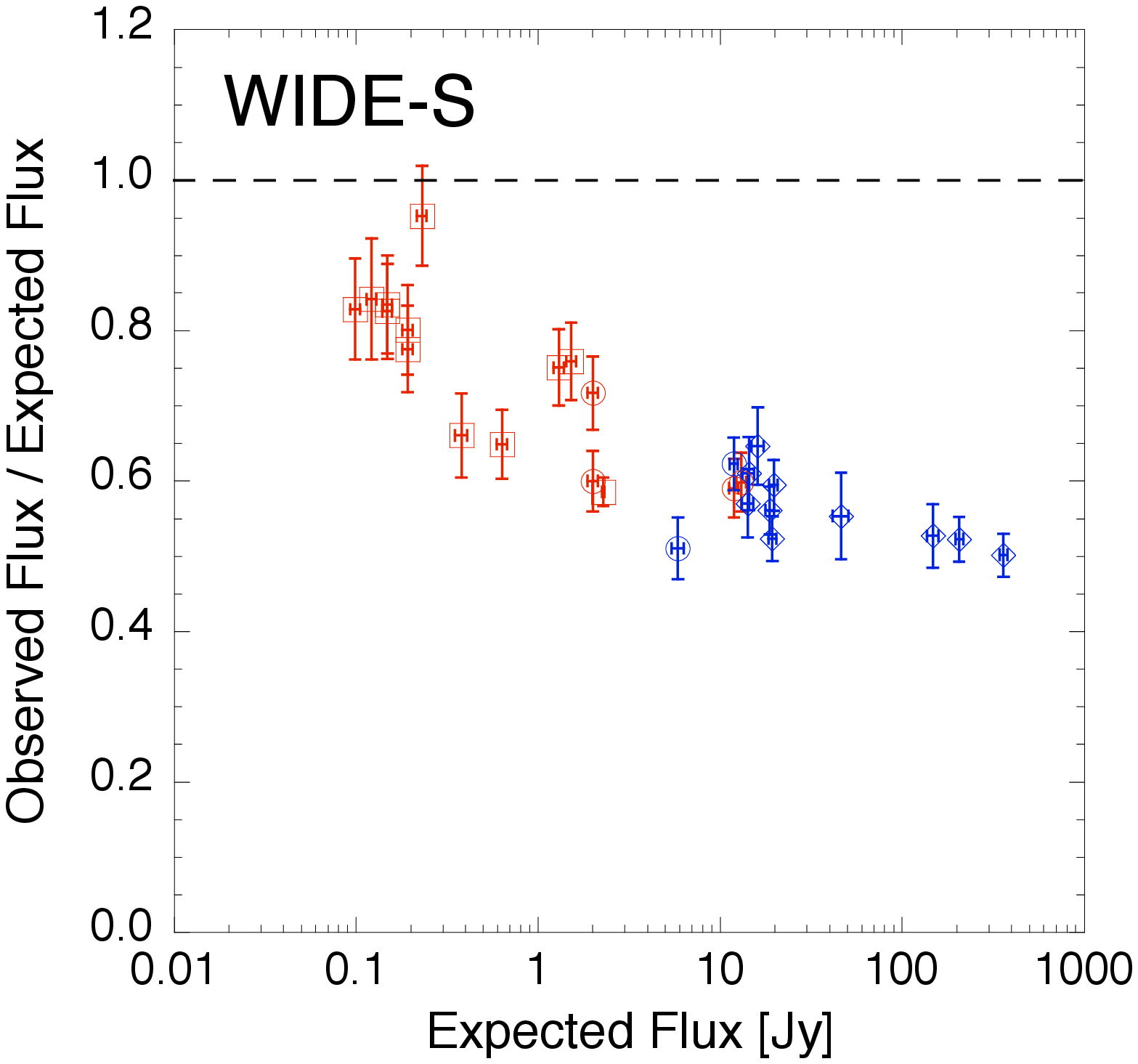}
    \FigureFile(80mm,80mm){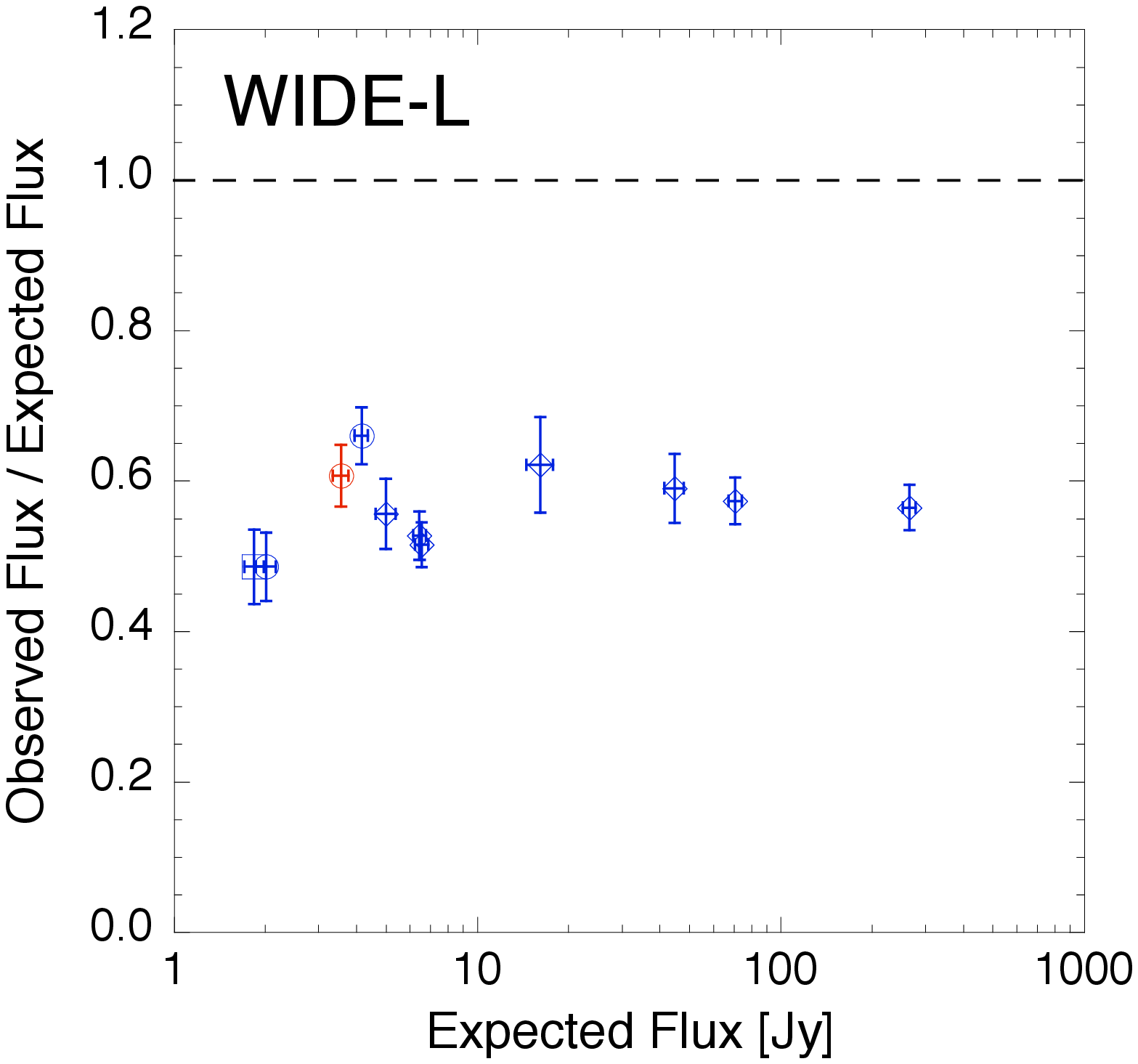} \quad
    \FigureFile(80mm,80mm){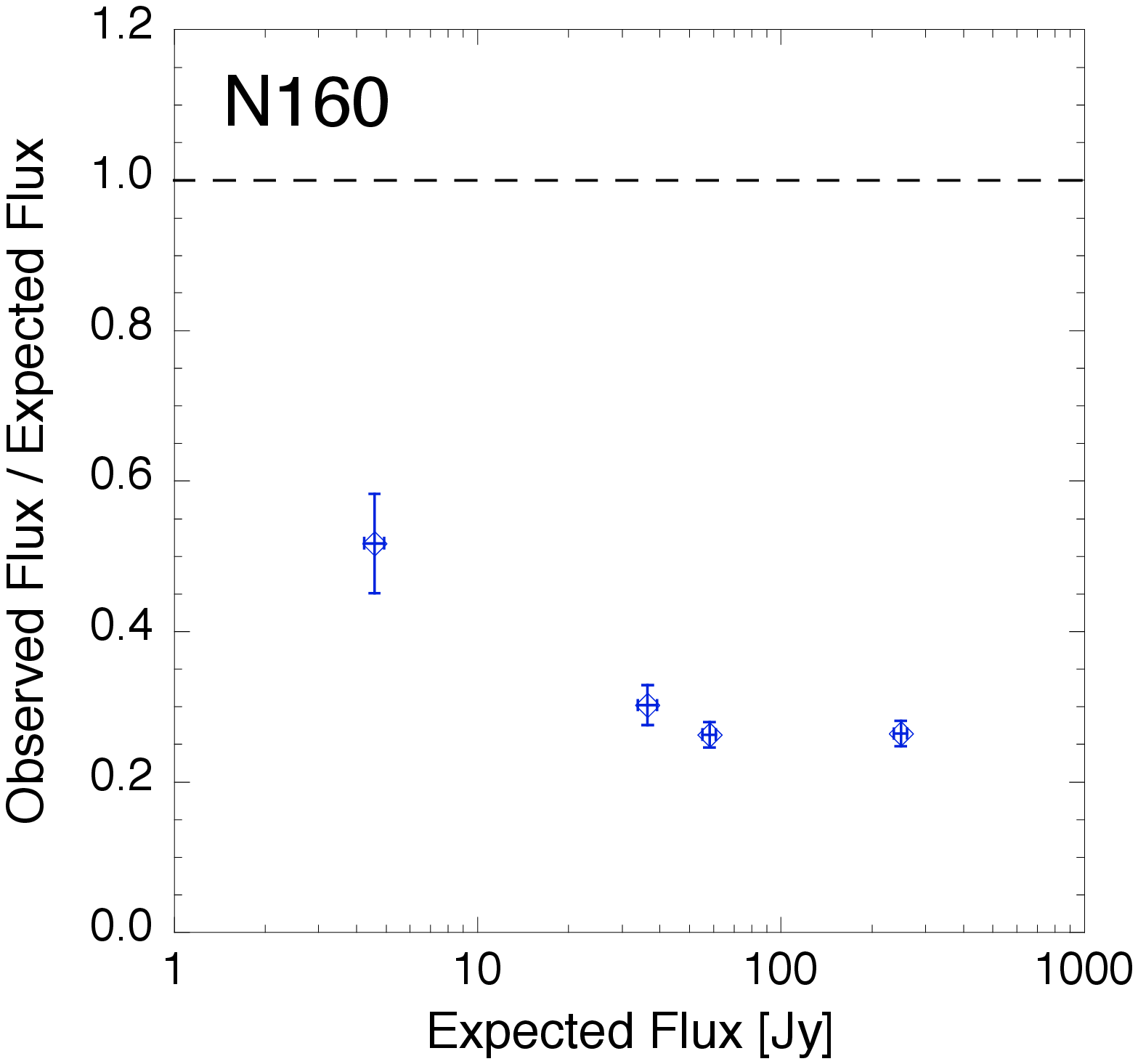}
  \end{center}
  \caption{
    The observed-to-expected flux ratio as a function of the expected flux. 
    The error bars on the y-axis represent the combined uncertainty of measurement errors and model uncertainties. 
    The red and blue symbols denote stars and asteroids, respectively. 
    The squares, circles, and diamonds are the data with reset intervals of 2, 1, and 0.5 sec, respectively. 
    The dashed lines indicate a linear relation.
    }\label{fig:dpraw}
\end{figure}

In order to derive the calibration factor for the point-source photometry, 
we compared the observed fluxes obtained by the aperture photometry with their model prediction. 
Figure~\ref{fig:dpraw} shows the observed-to-expected flux ratios for 
the calibration standards as a function of the expected flux. 
Two asteroids (Germania and Iris) are excluded from the plot 
because of their poor model accuracies.
The uncertainty on the flux ratios includes both photometric error and 
the expected model flux of the calibration standard. 
As shown in Fig.~\ref{fig:dpraw}, even after the aperture correction, 
the observed fluxes are always lower than the expected fluxes. 
This means that the sensitivity to the point-sources is lower than 
that to the diffuse sources. 
In addition, the deviation seems to have flux-dependency. 
A plausible cause of this disagreement is 
the slow transient response of the Ge:Ga detectors. 
Prior to the launch of AKARI, we had measured the slow response 
systematically under various photon flux conditions 
and established an empirical model (Kaneda et al. 2002). 
According to the model, the slow response should depend on the 
total photo-current (background plus signal). 

\begin{figure}
  \begin{center}
    \FigureFile(80mm,80mm){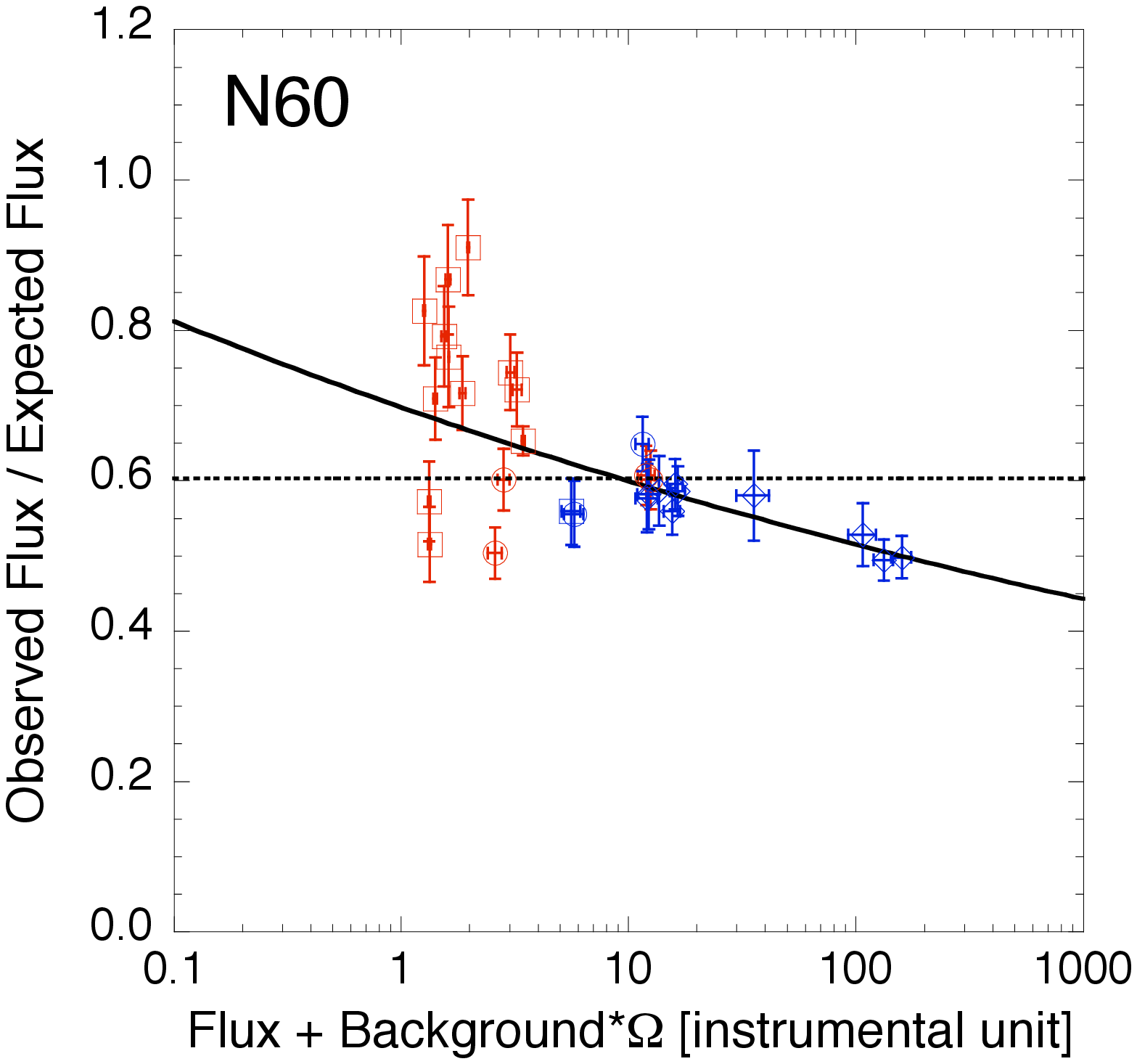} \quad
    \FigureFile(80mm,80mm){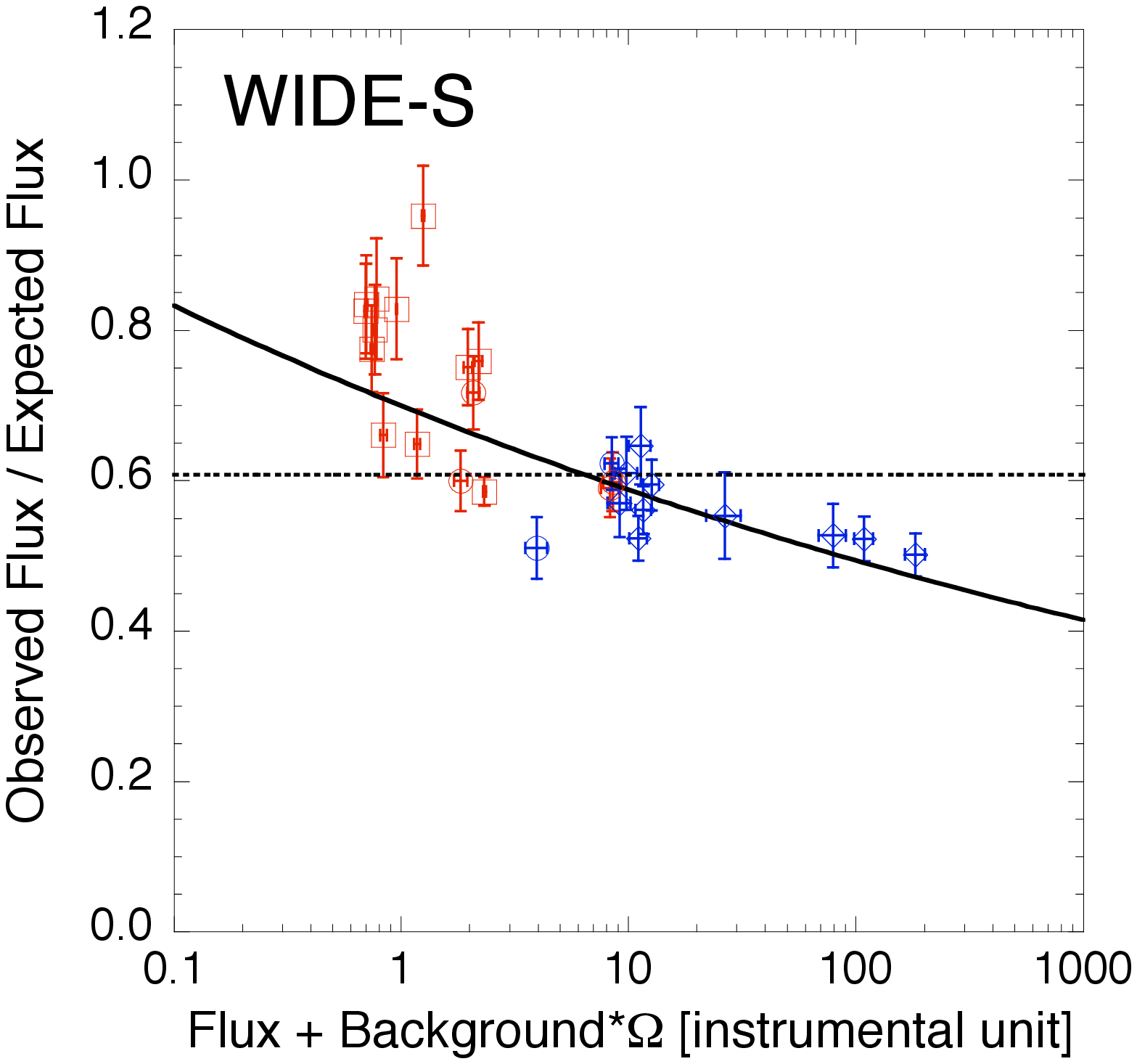}
    \FigureFile(80mm,80mm){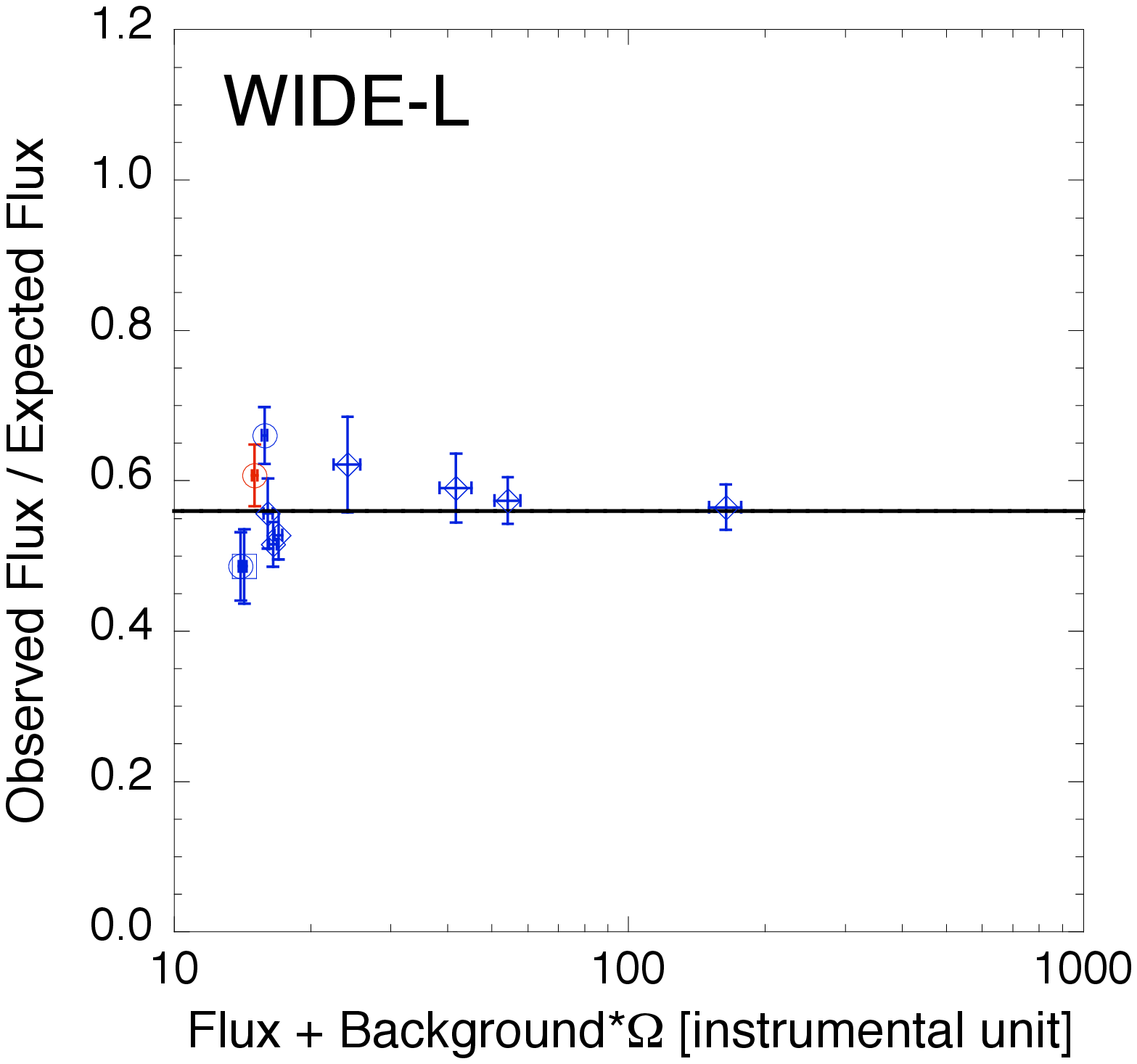} \quad
    \FigureFile(80mm,80mm){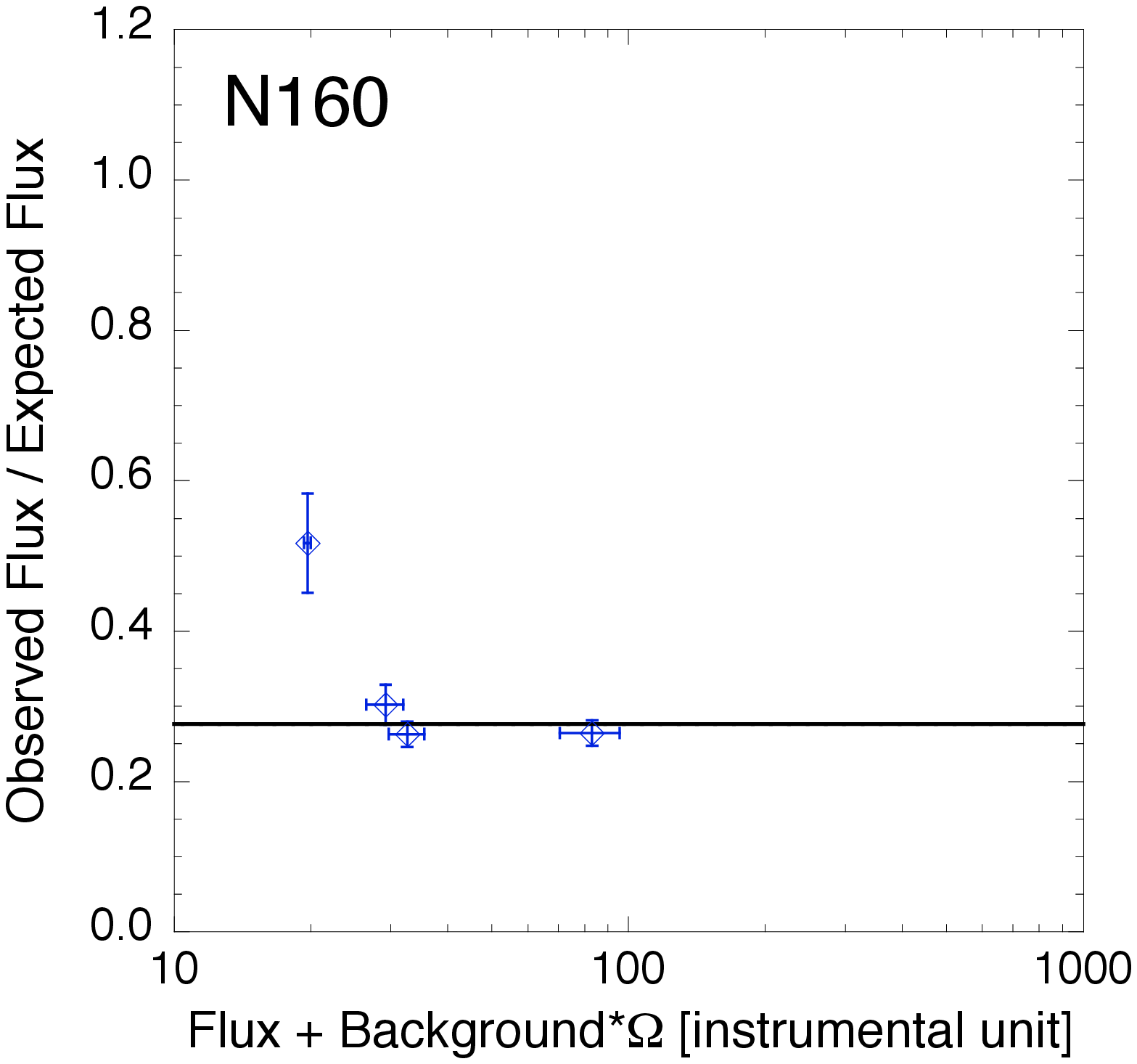}
  \end{center}
  \caption{
    The observed-to-expected flux ratio as a function of 
    the total observed flux including the background flux and the dark current.
    The color and symbols are the same as those in Fig.~\ref{fig:dpraw}. 
    The solid lines for the SW bands are the results of the power-law fit; 
    $y=0.698\times x^{-0.0659}$ ({\it N60}), 
    $y=0.700\times x^{-0.0757}$ ({\it WIDE-S}). 
    The dashed lines for the SW bands and the solid lines for the LW bands are the weighted average; 
    $y=0.560$ ({\it WIDE-L}), $y=0.277$ ({\it N160}). 
    }\label{fig:dp}
\end{figure}

Figure~\ref{fig:dp} shows the observed-to-expected flux ratios, 
as a function of the total observed flux including the background. 
The observed background used here is the sum of 
the background sky brightness and the detector dark current. 
In order to convert the units of the background from MJy~$\rm{sr^{-1}}$ to Jy, 
the beam solid angle derived from the PSF, 
which is summarized in Tab.~\ref{tab:bsa}, is multiplied. 

As described in Sec.~2, the reset interval was changed depending on the source flux. 
In Fig.~\ref{fig:dp}, different symbols denote data with different reset intervals. 
Because no systematic differences were seen, we deal with these data equally. 

In the SW bands, the background flux is so small that 
the total flux is dominated by the source flux. 
We can see a clear trend, the higher the total flux, the smaller the ratio. 
For bright sources, the ratio is $\sim$0.5,
while for faint sources, the ratio is $\sim$0.8, 
although they show large scatter. 
Solid lines in Fig.~\ref{fig:dp} are the results of power-law fitting to the data. 
The fitting results give
\begin{eqnarray}
  \mbox{Ratio} &=& (0.698 \pm 0.015) \times (\mbox{Total Flux})^{(-0.0659 \pm 0.0089)} \quad \mbox{for the {\it N60} band,} \nonumber \\
  \mbox{Ratio} &=& (0.700 \pm 0.013) \times (\mbox{Total Flux})^{(-0.0757 \pm 0.0085)} \quad \mbox{for the {\it WIDE-S} band.} \nonumber
\end{eqnarray}
The deviations of the data from the power-law are reasonably small, 13.7\% and 12.7\% 
for the {\it N60} and {\it WIDE-S} bands, respectively. 
If we assume that the ratio is constant, the weighted means are 
$0.603 \pm 0.007$ and $0.607 \pm 0.008$ 
for the {\it N60} and {\it WIDE-S} bands, respectively, 
and the deviation increases to approximately 20\% for both bands. 
Therefore, we conclude that there is a significant trend in the ratio with the total flux 
due to the slow transient response of the Ge:Ga detectors, 
and adopted the fitting results as the flux-dependent calibration factors. 

In contrast to the case of the SW bands, 
the total fluxes in the LW bands are dominated by the offset light. 
Therefore, the total flux range is limited small, 
and the flux ratio does not show a clear flux dependence. 
Unfortunately for the {\it N160} band, a few data points does not give 
a firm conclusion but can suggest a trend as well as the {\it WIDE-L} band. 
No dependence of the flux ratio on the small flux range has been expected from 
an empirical model of stressed Ge:Ga photoconductor (Kaneda et al. 2002). 
The weighted means of the ratio in Fig.~\ref{fig:dp} are $0.560 \pm 0.011$ and $0.277 \pm 0.011$ 
for the {\it WIDE-L} and {\it N160} bands, respectively. 
The results show that the LW detectors have much slower response than the SW detectors 
despite that it uses offset light to improve the slow transient response. 
The deviations of the data from the weighted means are 9.97\% and 50.5\% 
for the {\it WIDE-L} and {\it N160} bands, respectively. 
The calibration accuracy of the {\it N160} band are much lower 
than the other bands because of the small number of samples.

\subsection{Validity of the Calibration}

In this subsection, we perform various checks on the calibrations, 
such as comparison with previous missions, repeatability, 
the effects of scan speed, and observation mode. 
The results of the aperture photometry after applying 
the calibration factors derived above are listed in Tab.~\ref{tab:obsresult}.

\subsubsection{Comparison with Previous Missions}

In order to check the validity of the flux calibration, 
we observed the 11 galaxies listed in Tab.~\ref{tab:obslog}, and 
compared their AKARI fluxes with those from IRAS and ISO. 

\begin{figure}
  \begin{center}
    \FigureFile(80mm,80mm){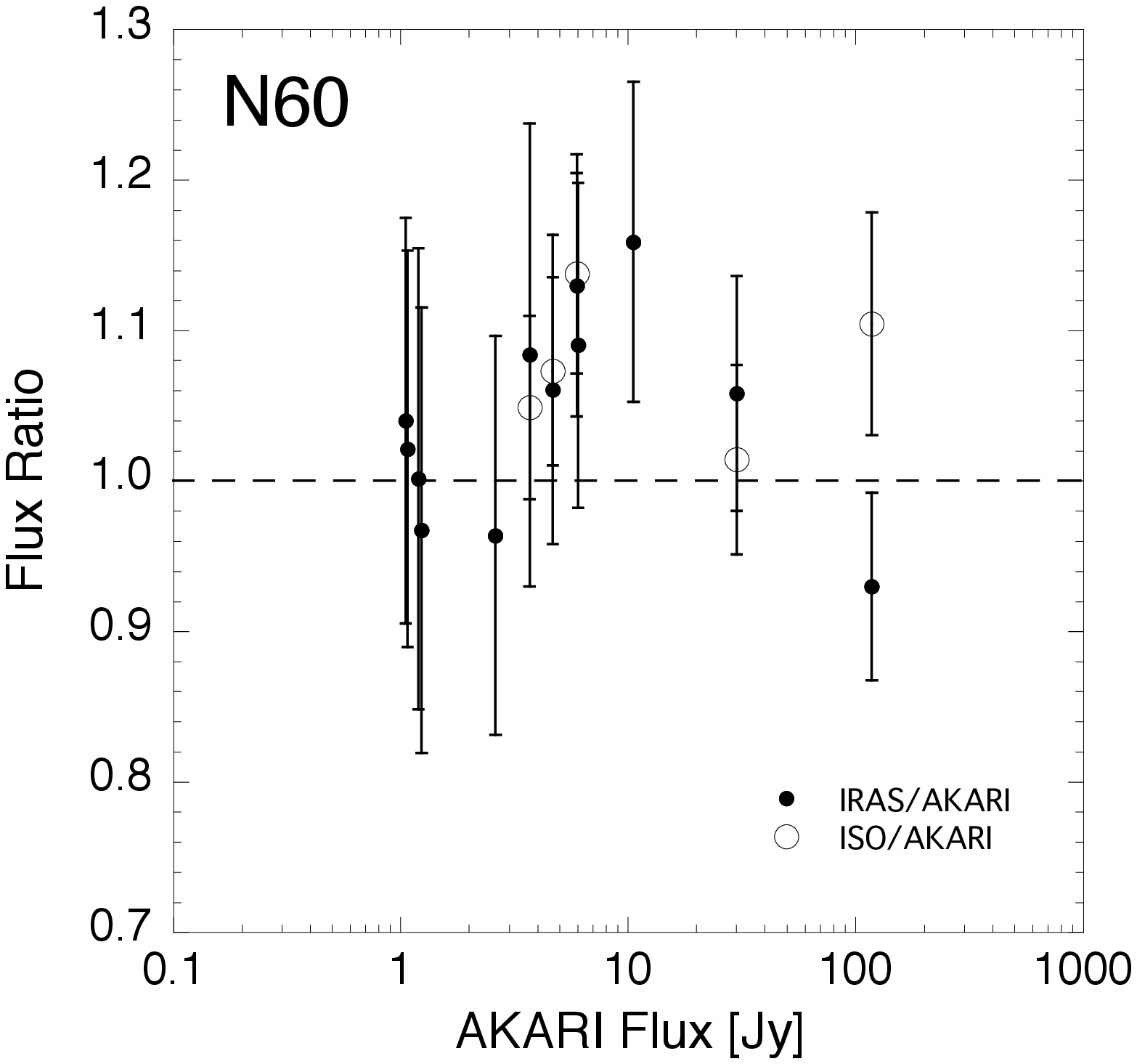} \quad
    \FigureFile(80mm,80mm){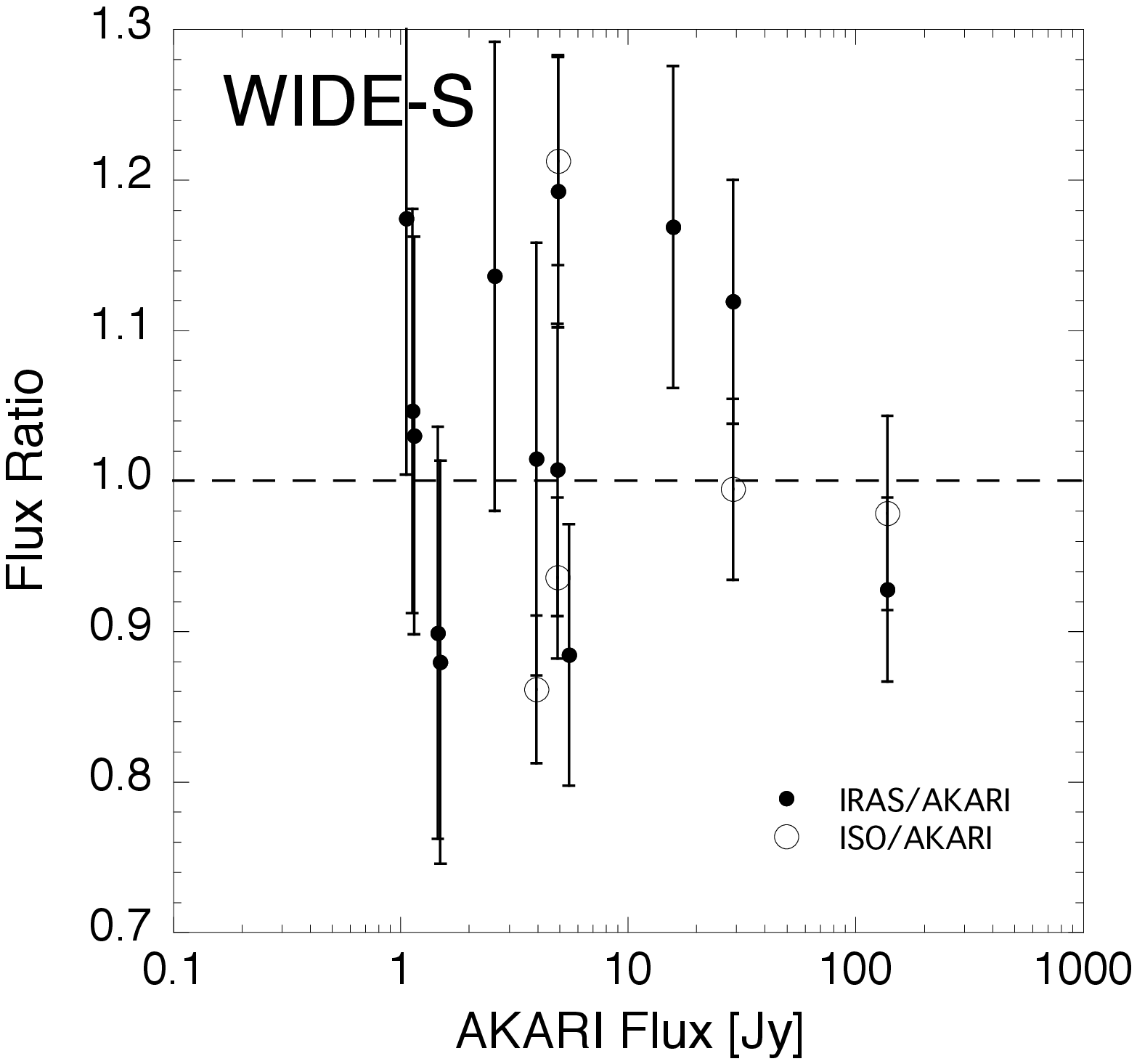}
  \end{center}
  \caption{
    The comparisons of AKARI flux with IRAS and ISO. 
    The filled circles represent the flux ratio of AKARI-to-IRAS, 
    while the opened circles represent the flux ratio of AKARI-to-ISO/ISPHOTO. 
    The plots were not displayed for the {\it WIDE-L} and {\it N160} bands, 
    because there were only two comparable data points. 
    }\label{fig:galhikaku}
\end{figure}

All 11 galaxies were detected by IRAS at both 60 and 100~$\rm{\mu m}$ 
(IRAS Faint Source Catalog; Moshir et al. 1992). 
The predictions for the 65 and 90~$\rm{\mu m}$ fluxes 
based on the IRAS measurements were calculated as follows. 
First, we apply color corrections to the IRAS measurements 
(Beichman et al. 1988) by assuming a power-law spectrum. 
Then we interpolate the IRAS 60 and 100~$\rm{\mu m}$ measurements to the 
center wavelength of the AKARI bands, 65 and 90~$\rm{\mu m}$. 
Finally, we apply color corrections to the AKARI measurements 
by assuming the same power law spectrum as that determined by the IRAS measurements. 
Figure~\ref{fig:galhikaku} shows the IRAS-to-AKARI flux ratios as 
a function of the AKARI flux. 
The average ratios of IRAS-to-AKARI measurements are 
$1.04\pm0.03$ and $1.03\pm0.03$ for the {\it N60} and {\it WIDE-S} bands, respectively. 
This result implies that the absolute calibration of both instruments 
consistent to each other within the uncertainties. 

Among our 11 samples, 5 famous galaxies (Arp 220, Mrk 231, IRAS 20100$-$4156, 
IRAS 15250$+$3609, IRAS 03158$+$4227) were observed by the ISO/ISOPHOT 
at 10 bands between 10 and 200~$\rm{\mu m}$ (Klaas et al. 2001). 
The wide wavelength coverage of ISO at longer than 100~$\rm{\mu m}$ enables us 
to compare them not only in the SW bands, but also in the LW bands. 
We calculate the average ratios of ISO/ISOPHOT-to-AKARI 
by using the same method as that of IRAS-to-AKARI, 
but by assuming a gray-body spectrum given by 
the ISO/ISOPHOT measurement (Klaas et al. 2001) rather than a power-law spectrum. 
The results of the comparison are also shown in Fig.~\ref{fig:galhikaku}. 
The ratios of ISO-to-AKARI are $1.07\pm0.03$, $0.98\pm0.03$, $0.97\pm0.13$, and $0.89\pm0.17$ 
for the {\it N60}, {\it WIDE-S}, {\it WIDE-L}, and {\it N160} bands, respectively. 
Although the number of sample is not statistically sufficient, 
the obtained values are equal to unity within the uncertainties.

\subsubsection{Repeatability}

The photometric repeatability was checked by the data of stars and galaxies 
that were observed twice; asteroids are not suitable for 
repeatability checks because of the time-variations in their fluxes. 
In the case of the SW bands, 
6 case studies (HR 5321, HR 5430, HR 872, HR 1208, IRAS 08201$+$2801, 
and IRAS 08474$+$1813) are available. Except for HR 1208, 
two measurements in each case agree to within 10\% and 3\% 
for the {\it N60} and {\it WIDE-S} bands, respectively. 
As for the {\it WIDE-L} and {\it N160} bands, 
since there is no comparable data, 
we could not check the photometric repeatability.

\subsubsection{Scan Speed}

In order to examine the dependence of the calibration factor on the scan speed, 
three calibrators (Vesta, Europa, and Neptune) were observed 
at scan speeds of both \timeform{8''}~$\rm{s^{-1}}$ and \timeform{15''}~$\rm{s^{-1}}$. 
These data are listed in Tab.~\ref{tab:obsresult}. 
The observed fluxes for the \timeform{15''}~$\rm{s^{-1}}$ data 
are $\sim$10\% fainter than those for the \timeform{8''}~$\rm{s^{-1}}$ data in all bands. 
This result implies contribution of the slow transient response of the Ge:Ga detector. 
However, the number of sample is statistically insufficient to give a quantitative conclusion.

\subsubsection{Observation Mode}

The calibration factors derived here are determined from observations 
by using AOT FIS01 (two round-trip scans) with a \timeform{70''} shift length, 
but should also apply directly to data taken with a \timeform{240''} shift length 
or with AOT FIS02 (single round-trip scan), 
because all of the modes conduct the same detector operation. 

In order to evaluate the influence by the difference of the observation mode, 
we checked data for the far-infrared deep survey observed by 
the AKARI mission program teams FBSEP (Shirahata et al. 2009, in prep.). 
They observed $\sim$2~${\rm deg}^2$ area with both FIS01 and FIS02 
at scan speeds of \timeform{15''}~$\rm{s^{-1}}$, 
and detected 8, 126, 4, and 1 galaxies in the {\it N60}, {\it WIDE-S}, 
{\it WIDE-L}, and {\it N160} bands, respectively. 
The obtained fluxes are consisted with each other within 
3\%, 1\%, 6\%, and 10\% for the {\it N60}, {\it WIDE-S}, 
{\it WIDE-L}, and {\it N160} bands, respectively, 
which are within the uncertainties. 
Therefore, we concluded that there are no significant differences 
originated in the observation mode.

\subsection{Application to the Data Reduction}

We note that our photometry is performed to a point-source with a fixed sky annulus; 
\timeform{2.3'}--\timeform{3.3'} in radius for the SW bands and 
\timeform{3.0'}--\timeform{4.0'} in radius for the LW bands. 
Therefore, the calibration factors presented here should only be applied 
to measurements carried out with the same parameters. 
Photometry using a different aperture size is possible 
via the application of an aperture correction factor (Tab.~\ref{tab:cog}), 
if the target is a point-source. If the target is an extended source, 
proper treatment for the slow response is necessary. 

Table~\ref{tab:abscal} summarizes the calibration factors for point-source photometry 
in the FIS Slow-Scan observations. In order to measure the point-source fluxes, 
the signals integrated within the aperture must be divided by the calibration factor 
to correct the slow transient response. 
For the point-sources brighter than 0.4, 0.1, 2.0, and 1.5~Jy 
in the {\it N60}, {\it WIDE-S}, {\it WIDE-L}, and {\it N160} bands, respectively, 
uncertainties of the flux calibrations are larger than the measurement errors. 

The FIS photometric flux is defined for a $\nu F_{\nu}=$~const. spectrum 
at the center wavelength of each band. 
This definition was first adopted by IRAS and 
used by several infrared astronomical satellites such as 
COBE and ISO. Also, the Spitzer/IRAC and IRS instruments use 
the same convention, while the Spitzer/MIPS uses a 10000~K 
black-body as its reference spectrum. 
The flux obtained using these derived calibration factors is 
not the actual flux, but a 'quoted' flux. 
Therefore, in order to obtain the monochromatic flux at the band center wavelength, 
we should apply a color correction depending on the SED of the target source. 
A color correction factor, $K$, is defined as 
$K = \Delta \nu_{\rm{SED}} / \Delta \nu_{\rm{flat}} = F_{\rm{obtained}} / F_{\rm{real}}$, 
where $\Delta \nu$ is the effective band width. 
Color correction factors assuming a gray-body spectrum and a power-law spectrum 
are shown in Tab.~\ref{tab:color}.

\section{Summary}

We performed flux calibrations for the FIS Slow-Scan observations, 
based on the measurements of stars and asteroids. 
We described in detail the data reduction and aperture photometric procedures 
that we used for the calibration sources. 
The expected fluxes of the calibration sources are lying in a wide 
flux range from 0.1~Jy to 400~Jy. 
There was a systematic flux dependence between the observed flux 
and the expected flux of the calibration sources, 
which was attributed to the slow transient response of the Ge:Ga detector. 
The calibration accuracies for the point-sources were estimated to be 
14\%, 13\%, 10\%, and 50\% for the {\it N60}, {\it WIDE-S}, 
{\it WIDE-L}, and {\it N160} bands, respectively. 
The calibration scheme described in this paper will be useful for future missions, 
such as Herschel and SPICA. 

\bigskip

AKARI is a JAXA project with the participation of ESA. 
We thank all of the members of the AKARI project for their continuous help and support. 
We would also express our gratitude to the FIS hardware and software team for their devoted work 
in developing the FIS instruments and its data reduction tool kit. 
The FIS was developed in collaboration with the ISAS/JAXA, Nagoya University, the University of Tokyo, 
the National Institute of Information and Communications Technology (NICT), 
the National Astronomical Observatory of Japan (NAOJ), and other research institutes. 
We are grateful to S.~Oyabu for help with the development of the reduction tools. 
We thank to M.~Fukagawa and T.~Hirao for valuable discussions. 
We thank to W.-S.~Jeong for providing the data of the optical simulation model of PSF, and 
thank to R.~Moreno for providing flux model of Neptune.  
This work is partly supported by JSPS grants (16204013 and 19540250). 
M.~Cohen thanks ISAS/JAXA for supporting a visit to Tokyo to participate in this work. 
S.~Hasegawa was supported by Space Plasma Laboratory, ISAS/JAXA.

\newpage

\begin{table}
  \begin{center}
    \caption{Specification for the FIS instrument in photometry mode.}
    \label{tab:fis}
    \begin{tabular}{lccccl}
      \hline
      Band name            & {\it N60}   & {\it WIDE-S} & {\it WIDE-L} & {\it N160}  &  \\ \hline
      Center wavelength    & 65          & 90           & 140          & 160         & [$\rm{\mu m}$] \\
      Wavelength range     & 50--80      & 60--110      & 110--180     & 140--180    & [$\rm{\mu m}$] \\ \hline
      Array format         & 20$\times$2 & 20$\times$3  & 15$\times$3  & 15$\times$2 & [pixels] \\
      Pixel scale$^{\ast}$   & \multicolumn{2}{c}{26.8$\times$26.8} & \multicolumn{2}{c}{44.2$\times$44.2} & [$\rm{arcsec}^2$]  \\
      Pixel pitch$^{\ast}$  & \multicolumn{2}{c}{29.5$\times$29.5} & \multicolumn{2}{c}{49.1$\times$49.1} & [$\rm{arcsec}^2$]  \\ \hline
      Detector device      & \multicolumn{2}{c}{Monolithic Ge:Ga Array} & \multicolumn{2}{c}{Stressed Ge:Ga Array} & \\ \hline
      \\
      \multicolumn{6}{@{}l@{}}{\hbox to 0pt{\parbox{180mm}{\footnotesize
            Notes. 
            \par\noindent
            \footnotemark[$*$] at the center of the field of view. 
          }\hss}}
    \end{tabular}
  \end{center}
\end{table}

\begin{table}
  \begin{center}
    \caption{Observation log and expected flux.}
    \label{tab:obslog}
    {\scriptsize
      \begin{tabular}{lllcrrrrr}
        \hline
Target name           & \multicolumn{2}{c}{Observation}                & & \multicolumn{5}{c}{Expected flux$^{\dagger}$}                   \\ \cline{2-3} \cline{5-9}
                      & Date                & AOT & &      {\it N60} &      {\it WIDE-S} &      {\it WIDE-L} &    {\it N160} & Accuracy \\
                      &                     & parameter$^{\ast}$  & &     [Jy] &     [Jy] &     [Jy] &    [Jy] &     [\%] \\ \hline

HR 5826               & 2006/04/21 20:38:35 & 2.0;8;70   & &         0.338  &          0.237 &    0.064 &   0.052 &  6   \\
HR 5321               & 2006/04/22 06:31:28 & 2.0;8;70   & &         0.276  &          0.193 &    0.053 &   0.043 &  6   \\
HR 5321 (2)           & 2006/04/22 11:28:30 & 2.0;8;70   & &         0.276  &          0.193 &    0.053 &   0.043 &  6   \\
HR 5430               & 2006/04/28 04:27:17 & 2.0;8;70   & &         0.543  &          0.380 &    0.104 &   0.084 &  6   \\
HR 5430 (2)           & 2006/04/28 06:06:18 & 2.0;8;70   & &         0.543  &          0.380 &    0.104 &   0.084 &  6   \\
HR 1208               & 2006/04/29 15:57:40 & 1.0;8;70   & &         2.864  &          2.006 &    0.545 &   0.441 &  6   \\
HR 872                & 2006/04/30 00:11:43 & 2.0;8;70   & &         0.214  &          0.150 &    0.041 &   0.033 &  6   \\
HR 872 (2)            & 2006/04/30 01:50:45 & 2.0;8;70   & &         0.214  &          0.150 &    0.041 &   0.033 &  6   \\
HR 1208 (2)           & 2006/05/02 01:43:26 & 1.0;8;70   & &         2.864  &          2.006 &    0.545 &   0.441 &  6   \\
Alpha CMa             & 2006/10/07 18:28:06 & 2.0;8;70   & &         3.290  &    {\bf 2.293} &    0.616 &   0.497 &  1.47 \\
Alpha Boo             & 2007/01/15 00:02:26 & 1.0;8;70   & &   {\bf 18.689} &   {\bf 13.089} &    3.558 &   2.879 &  6   \\
Alpha Tau             & 2007/02/28 14:18:57 & 1.0;8;70   & &   {\bf 17.042} &   {\bf 11.939} &    3.249 &   2.630 &  6   \\
HD 216386             & 2007/06/03 01:05:46 & 2.0;8;70   & &         2.177  &          1.524 &    0.414 &   0.335 &  6   \\
HD 98118              & 2007/06/10 01:17:14 & 2.0;8;70   & &         0.330  &          0.232 &    0.063 &   0.051 &  6   \\
HD 222643             & 2007/06/11 01:23:08 & 2.0;8;70   & &         0.142  &          0.099 &    0.027 &   0.022 &  6   \\
HD 224935             & 2007/06/20 00:48:29 & 2.0;8;70   & &         1.869  &          1.309 &    0.355 &   0.288 &  6   \\
HD 053501             & 2007/07/13 02:52:02 & 2.0;8;70   & &         0.175  &          0.122 &    0.033 &   0.027 &  6   \\
HD 92305              & 2007/08/23 12:12:43 & 2.0;8;70   & &         0.906  &          0.636 &    0.173 &   0.140 &  6   \\ \hline

241 Germania          & 2006/04/27 15:44:31 & 0.5;8;70   & &         8.958  &    {\bf 6.932} &    2.356 &   1.940 & 12.5 \\
241 Germania (2)      & 2006/04/27 23:59:37 & 0.5;8;70   & &         7.813  &    {\bf 6.064} &    2.073 &   1.707 & 12.5 \\
6 Hebe                & 2006/04/30 03:07:09 & 0.5;8;70   & &   {\bf 25.258} &   {\bf 19.382} &    6.469 &   5.313 &  5   \\
6 Hebe (2)            & 2006/05/01 00:34:26 & 0.5;8;70   & &   {\bf 25.681} &   {\bf 19.699} &    6.570 &   5.396 &  5   \\
511 Davida            & 2006/05/02 22:50:20 & 0.5;8;70   & &   {\bf 18.394} &   {\bf 14.387} &    4.999 &   4.127 &  7.5 \\
511 Davida (2)        & 2006/05/03 12:02:32 & 0.5;8;70   & &   {\bf 18.185} &   {\bf 14.214} &    4.933 &   4.071 &  7.5 \\
7 Iris                & 2006/08/01 18:19:43 & 0.5;8;70   & &        56.355  &        42.927  &   14.103 &  11.554 & 20   \\
2 Pallas              & 2006/09/27 06:20:31 & 0.5;8;70   & &   {\bf 59.254} &   {\bf 46.375} &   16.142 &  13.329 & 10   \\
1 Ceres               & 2006/11/08 14:58:11 & 0.5;8;70   & &  {\bf 264.848} &  {\bf 206.126} &   {\bf 70.786} &  {\bf 58.327} &  5   \\
93 Minerva            & 2006/11/20 00:42:13 & 1.0;8;70   & &         7.551  &    {\bf 5.873} &    2.017 &   1.662 &  7.5 \\
65 Cybele             & 2006/12/28 00:16:17 & 1.0;8;70   & &   {\bf 15.192} &   {\bf 11.905} &    4.155 &   3.431 &  5   \\
4 Vesta               & 2007/02/23 22:33:11 & 0.5;8;70   & &  {\bf 200.598} &  {\bf 147.871} &   {\bf 44.748} &  36.486 &  7.5 \\
4 Vesta (2)           & 2007/02/24 00:12:31 & 0.5;15;70  & &  {\bf 202.519} &  {\bf 149.228} &   {\bf 45.113} &  36.778 &  7.5 \\
52 Europa             & 2007/04/14 23:08:31 & 0.5;8;70   & &   {\bf 24.150} &   {\bf 18.807} &    6.467 &   5.328 &  5   \\
52 Europa (2)         & 2007/04/15 22:19:51 & 0.5;15;70  & &   {\bf 24.328} &   {\bf 18.941} &    6.511 &   5.364 &  5   \\
Neptune               & 2007/05/13 01:22:57 & 0.5;8;70   & &  {\bf 315.942} &  {\bf 361.867} &  {\bf 265.605} & {\bf 248.897} &  5   \\
Neptune (2)           & 2007/05/13 19:36:26 & 0.5;15;70  & &  {\bf 316.215} &  {\bf 362.171} &  {\bf 265.833} & {\bf 249.113} &  5   \\
47 Aglaja             & 2007/06/26 01:48:04 & 2.0;8;70   & &         7.008  &         5.423  &    1.844 &   1.518 &  7.5 \\
511 Davida (3)        & 2007/07/20 03:36:26 & 0.5;8;70   & &   {\bf 20.743} &   {\bf 16.175} &    5.576 &   4.592 &  7.5 \\ \hline

IRAS 08201$+$2801     & 2006/04/22 01:17:21 & 0.5;8;70   & &    --  &             -- &       -- &      -- &   -- \\
IRAS 08201$+$2801 (2) & 2006/04/22 02:56:21 & 0.5;8;70   & &    --  &             -- &       -- &      -- &   -- \\
IRAS 08591$+$5248     & 2006/04/22 04:42:28 & 0.5;8;70   & &    --  &             -- &       -- &      -- &   -- \\
IRAS 08572$+$3915     & 2006/04/26 01:03:51 & 0.5;8;70   & &    --  &     $\bigcirc$ &       -- &      -- &   -- \\
IRAS 08474$+$1813     & 2006/04/30 03:59:18 & 0.5;8;70   & &    --  &             -- &       -- &      -- &   -- \\
IRAS 08474$+$1813 (2) & 2006/04/30 05:38:20 & 0.5;8;70   & &    --  &             -- &       -- &      -- &   -- \\
Arp 220               & 2006/08/06 06:45:02 & 0.5;8;70   & & $\bigcirc$ &     $\bigcirc$ & $\bigcirc$ & $\bigcirc$ &   -- \\
Mrk 231               & 2006/12/01 19:20:05 & 0.5;8;70   & & $\bigcirc$ &     $\bigcirc$ &       -- &      -- &   -- \\
IRAS 20100$-$4156     & 2007/04/16 23:17:36 & 1.0;8;70   & &    --  &     $\bigcirc$ &       -- &      -- &   -- \\
UGC 05101             & 2007/04/22 12:44:18 & 0.5;8;70   & & $\bigcirc$ &     $\bigcirc$ &       -- &      -- &   -- \\
IRAS 00188$-$0856     & 2007/06/23 01:45:19 & 2.0;8;70   & &    --  &     $\bigcirc$ &       -- &      -- &   -- \\
IRAS 15250$+$3609     & 2007/07/28 03:16:26 & 2.0;8;70   & &    --  &     $\bigcirc$ &       -- &      -- &   -- \\
IRAS 03158$+$4227     & 2007/08/23 11:35:18 & 2.0;8;70   & &    --  &     $\bigcirc$ &       -- &      -- &   -- \\ \hline \\
\multicolumn{9}{@{}l@{}}{\hbox to 0pt{\parbox{180mm}{\footnotesize
      Notes. 
      \par\noindent
      \footnotemark[$*$] FIS01 AOT parameter : Reset interval [s]; scan speed [arcsec sec$^{-1}$]; shift length [arcsec].
      \par\noindent
      \footnotemark[$^{\dagger}$] Model calculation : Stars; Cohen et al. 1999, 2003a, 2003b, Asteroids; M$\rm{\ddot{u}}$ller \& Lagerros 1998, 2002, Neptune; Moreno 1998.
      \par\noindent
      The data designated with the bold-faced or $\bigcirc$ symbol is data used for the measurements of PSFs and the encircled energy function.
    }\hss}}
\end{tabular}
}
\end{center}
\end{table}

\begin{table}
  \begin{center}
    \caption{Gaussian fitting parameters for the PSFs.}
    \label{tab:psf}
    \begin{tabular}{lccccl}
      \hline
      Band name            & {\it N60}        & {\it WIDE-S}     & {\it WIDE-L}      & {\it N160}        &          \\ \hline
      Main component:      &                  &                  &                   &                   &          \\
      - Intensity (A)      & 79.03 $\pm$ 0.33 & 79.68 $\pm$ 0.23 &  76.01 $\pm$ 0.38 &  79.86 $\pm$ 0.30 & [\%]     \\
      - FWHM (based on $\sigma_1$)              & 32.05 $\pm$ 0.10 & 30.17 $\pm$ 0.08 &  40.85 $\pm$ 0.10 &  38.23 $\pm$ 0.15 & [arcsec] \\ \hline
      Sub component:       &                  &                  &                   &                   &          \\
      - Intensity (1$-$A)  & 20.97 $\pm$ 0.35 & 20.33 $\pm$ 0.24 &  23.99 $\pm$ 0.39 &  20.14 $\pm$ 0.30 & [\%]     \\
      - FWHM (based on $\sigma_2$)               & 77.51 $\pm$ 0.57 & 86.01 $\pm$ 0.53 & 118.98 $\pm$ 1.03 & 120.62 $\pm$ 0.51 & [arcsec] \\ \hline \\
      \multicolumn{6}{@{}l@{}}{\hbox to 0pt{\parbox{180mm}{\footnotesize
            Notes. 
            \par\noindent
            Two components Gaussian function: $ I(x) = A \exp{(-(x^2)/2\sigma_1^2)} + (1-A) \exp{(-(x^2)/2\sigma_2^2)} $;
            \par\noindent
            $A=$ Intensity of main component; $\sigma = {\rm FWHM}/2\sqrt{2 \ln 2}$.
          }\hss}}
    \end{tabular}
  \end{center}
\end{table}

\begin{table}
  \begin{center}
    \caption{Information of the ghost signal.}
    \label{tab:ghost}
    \begin{tabular}{lccl}
      \hline
      Detector          & SW {\small ({\it N60} $\leftrightarrow$ {\it WIDE-S})} & LW {\small ({\it WIDE-L} $\leftrightarrow$ {\it N160})} & \\ \hline
      Relative position\footnotemark[$*$]  &                        &                              &          \\
       - In-scan                           & 5.4                    & 3.6                          & [arcmin] \\
       - Cross-scan                        & 0.2                    & 2.0                          & [arcmin] \\
       - Distance                          & 5.4                    & 4.1                          & [arcmin] \\ \hline
      Intensity\footnotemark[$\dagger$]    & $\sim$1                & $\sim$10                     & [\%]     \\ \hline \\
      \multicolumn{4}{@{}l@{}}{\hbox to 0pt{\parbox{180mm}{\footnotesize
            Note. 
            \par\noindent
            The ghost signal appears in one array when a strong light enters the other array of the same detector. 
            \par\noindent
            \footnotemark[$*$] The relative position of the ghost image from the target source.
            \par\noindent
            \footnotemark[$\dagger$] The intensity of the ghost signal compared with the target signal observed in the other array.
          }\hss}}
    \end{tabular}
  \end{center}
\end{table}

\begin{table}
  \begin{center}
    \caption{Aperture correction factors.}
    \label{tab:cog}
    {\scriptsize
    \begin{tabular}{rcccc}
      \hline
    Aperture\footnotemark[$*$] &      {\it N60}      &     {\it WIDE-S}    &     {\it WIDE-L}    &     {\it N160}      \\ \hline
      5      &  0.030 $\pm$ 0.002  &  0.028 $\pm$ 0.002  &  0.015 $\pm$ 0.001  &  0.018 $\pm$ 0.001  \\
     10      &  0.112 $\pm$ 0.005  &  0.105 $\pm$ 0.006  &  0.056 $\pm$ 0.002  &  0.066 $\pm$ 0.003  \\
     15      &  0.225 $\pm$ 0.010  &  0.208 $\pm$ 0.010  &  0.116 $\pm$ 0.004  &  0.133 $\pm$ 0.009  \\
     20      &  0.340 $\pm$ 0.014  &  0.310 $\pm$ 0.014  &  0.187 $\pm$ 0.007  &  0.214 $\pm$ 0.018  \\
     25      &  0.440 $\pm$ 0.016  &  0.397 $\pm$ 0.016  &  0.264 $\pm$ 0.009  &  0.293 $\pm$ 0.025  \\
     30      &  0.524 $\pm$ 0.016  &  0.469 $\pm$ 0.017  &  0.337 $\pm$ 0.011  &  0.363 $\pm$ 0.026  \\
     35      &  0.596 $\pm$ 0.016  &  0.534 $\pm$ 0.016  &  0.402 $\pm$ 0.014  &  0.428 $\pm$ 0.026  \\
     40      &  0.659 $\pm$ 0.015  &  0.596 $\pm$ 0.016  &  0.458 $\pm$ 0.015  &  0.484 $\pm$ 0.026  \\
     45      &  0.715 $\pm$ 0.014  &  0.655 $\pm$ 0.015  &  0.507 $\pm$ 0.014  &  0.532 $\pm$ 0.027  \\
     50      &  0.762 $\pm$ 0.013  &  0.708 $\pm$ 0.014  &  0.554 $\pm$ 0.013  &  0.578 $\pm$ 0.028  \\
     55      &  0.799 $\pm$ 0.011  &  0.753 $\pm$ 0.014  &  0.599 $\pm$ 0.012  &  0.622 $\pm$ 0.028  \\
     60      &  0.826 $\pm$ 0.010  &  0.789 $\pm$ 0.013  &  0.644 $\pm$ 0.011  &  0.664 $\pm$ 0.028  \\
     65      &  0.847 $\pm$ 0.009  &  0.818 $\pm$ 0.013  &  0.689 $\pm$ 0.010  &  0.708 $\pm$ 0.028  \\
     70      &  0.865 $\pm$ 0.008  &  0.844 $\pm$ 0.013  &  0.734 $\pm$ 0.010  &  0.751 $\pm$ 0.028  \\
     75      &  0.882 $\pm$ 0.007  &  0.865 $\pm$ 0.012  &  0.778 $\pm$ 0.010  &  0.795 $\pm$ 0.028  \\
     80      &  0.897 $\pm$ 0.007  &  0.884 $\pm$ 0.012  &  0.817 $\pm$ 0.011  &  0.837 $\pm$ 0.028  \\
     85      &  0.911 $\pm$ 0.006  &  0.900 $\pm$ 0.011  &  0.852 $\pm$ 0.012  &  0.875 $\pm$ 0.028  \\
     90      &  0.923 $\pm$ 0.006  &  0.913 $\pm$ 0.010  &  0.881 $\pm$ 0.012  &  0.910 $\pm$ 0.027  \\
     95      &  0.934 $\pm$ 0.005  &  0.926 $\pm$ 0.009  &  0.904 $\pm$ 0.012  &  0.938 $\pm$ 0.026  \\
    100      &  0.944 $\pm$ 0.005  &  0.937 $\pm$ 0.008  &  0.922 $\pm$ 0.012  &  0.958 $\pm$ 0.024  \\
    105      &  0.954 $\pm$ 0.004  &  0.947 $\pm$ 0.007  &  0.936 $\pm$ 0.011  &  0.973 $\pm$ 0.022  \\
    110      &  0.963 $\pm$ 0.004  &  0.957 $\pm$ 0.006  &  0.948 $\pm$ 0.010  &  0.983 $\pm$ 0.019  \\
    115      &  0.971 $\pm$ 0.003  &  0.964 $\pm$ 0.005  &  0.956 $\pm$ 0.009  &  0.991 $\pm$ 0.017  \\
    120      &  0.977 $\pm$ 0.003  &  0.971 $\pm$ 0.005  &  0.963 $\pm$ 0.008  &  0.996 $\pm$ 0.015  \\
    125      &  0.982 $\pm$ 0.002  &  0.976 $\pm$ 0.004  &  0.969 $\pm$ 0.007  &  1.001 $\pm$ 0.013  \\
    130      &  0.987 $\pm$ 0.002  &  0.983 $\pm$ 0.003  &  0.975 $\pm$ 0.006  &  1.004 $\pm$ 0.012  \\
    135      &  0.991 $\pm$ 0.002  &  0.988 $\pm$ 0.002  &  0.979 $\pm$ 0.005  &  1.006 $\pm$ 0.010  \\
    140      &                     &                     &  0.983 $\pm$ 0.004  &  1.007 $\pm$ 0.009  \\
    145      &                     &                     &  0.986 $\pm$ 0.004  &  1.009 $\pm$ 0.008  \\
    150      &                     &                     &  0.989 $\pm$ 0.003  &  1.009 $\pm$ 0.007  \\
    155      &                     &                     &  0.991 $\pm$ 0.003  &  1.009 $\pm$ 0.007  \\
    160      &                     &                     &  0.992 $\pm$ 0.004  &  1.007 $\pm$ 0.007  \\
    165      &                     &                     &  0.994 $\pm$ 0.004  &  1.006 $\pm$ 0.007  \\
    170      &                     &                     &  0.995 $\pm$ 0.003  &  1.004 $\pm$ 0.006  \\
    175      &                     &                     &  0.996 $\pm$ 0.002  &  1.003 $\pm$ 0.005  \\
    180      &                     &                     &  0.998 $\pm$ 0.001  &  1.001 $\pm$ 0.003  \\ \hline \\
      \multicolumn{5}{@{}l@{}}{\hbox to 0pt{\parbox{180mm}{\footnotesize
            Notes. 
            \par\noindent
            \footnotemark[$*$] Aperture radius [arcsec]
          }\hss}}
    \end{tabular}
    }
  \end{center}
\end{table}

\begin{table}
  \begin{center}
    \caption{Beam solid angles.}
    \label{tab:bsa}
    \begin{tabular}{ll}
      \hline
      Band name     & Beam solid angle\footnotemark[$*$] [sr]\\ \hline
      {\it N60}     & $(4.06 \pm 0.10) \times 10^{-8}$ \\
      {\it WIDE-S}  & $(4.81 \pm 0.13) \times 10^{-8}$ \\
      {\it WIDE-L}  & $(9.92 \pm 0.19) \times 10^{-8}$ \\
      {\it N160}    & $(10.21 \pm 0.42) \times 10^{-8}$ \\ \hline \\
      \multicolumn{2}{@{}l@{}}{\hbox to 0pt{\parbox{180mm}{\footnotesize
            Note. 
            \par\noindent
            \footnotemark[$*$] The beam solid angles: $\pi r^2 \times \mbox{(Aperture correction factors at $r$})$; \\
            \quad $r=$ \timeform{40''} (for the {\it N60} and {\it WIDE-S} bands); \\
            \quad $r=$ \timeform{60''} (for the {\it WIDE-L} and {\it N160} bands). 
          }\hss}}
    \end{tabular}
  \end{center}
\end{table}

\begin{table}
  \caption{Results of observation.}\label{tab:obsresult}
  \begin{center}
    {\scriptsize
      \begin{tabular}{lrrrr}
      \hline
      Target name           & \multicolumn{4}{c}{Observed flux [Jy] }                                             \\ \cline{2-5}
                            & {\it N60}                  & {\it WIDE-S}               & {\it WIDE-L}         & {\it N160}           \\ \hline

      HR 5826               &   {\bf 0.351 $\pm$  0.013} &                --          &           --         &           --         \\
      HR 5321               &   {\bf 0.208 $\pm$  0.011} &   {\bf 0.209 $\pm$  0.007} &           --         &           --         \\
      HR 5321 (2)           &   {\bf 0.230 $\pm$  0.012} &   {\bf 0.216 $\pm$  0.008} &           --         &           --         \\
      HR 5430               &   {\bf 0.696 $\pm$  0.022} &                --          &           --         &           --         \\
      HR 5430 (2)           &   {\bf 0.634 $\pm$  0.020} &   {\bf 0.354 $\pm$  0.012} &           --         &           --         \\
      HR 1208               &   {\bf 2.199 $\pm$  0.064} &   {\bf 1.798 $\pm$  0.048} &           --         &           --         \\
      HR 872                &   {\bf 0.257 $\pm$  0.012} &   {\bf 0.172 $\pm$  0.007} &           --         &           --         \\
      HR 872 (2)            &                --          &   {\bf 0.174 $\pm$  0.007} &           --         &           --         \\
      HR 1208 (2)           &   {\bf 2.641 $\pm$  0.077} &   {\bf 2.172 $\pm$  0.058} &           --         &           --         \\
      Alpha CMa             &   {\bf 3.337 $\pm$  0.097} &   {\bf 2.045 $\pm$  0.056} &           --         &           --         \\
      Alpha Boo             &  {\bf 19.009 $\pm$  0.655} &  {\bf 13.163 $\pm$  0.403} &   {\bf 3.855 $\pm$  0.101} &           --         \\
      Alpha Tau             &  {\bf 17.452 $\pm$  0.598} &  {\bf 11.815 $\pm$  0.360} &           --         &           --         \\
      HD 216386             &   {\bf 2.430 $\pm$  0.070} &   {\bf 1.753 $\pm$  0.047} &           --         &           --         \\
      HD 98118              &   {\bf 0.450 $\pm$  0.014} &   {\bf 0.321 $\pm$  0.009} &           --         &           --         \\
      HD 222643             &   {\bf 0.161 $\pm$  0.007} &   {\bf 0.117 $\pm$  0.005} &           --         &           --         \\
      HD 224935             &   {\bf 2.143 $\pm$  0.062} &   {\bf 1.478 $\pm$  0.039} &           --         &           --         \\
      HD 053501             &                --          &   {\bf 0.144 $\pm$  0.008} &           --         &           --         \\
      HD 92305              &   {\bf 0.968 $\pm$  0.028} &   {\bf 0.596 $\pm$  0.018} &           --         &           --         \\ \hline

      241 Germania          &       10.249 $\pm$  0.329  &        7.795 $\pm$  0.226  &        2.806 $\pm$  0.109  &           --         \\
      241 Germania (2)      &       10.390 $\pm$  0.334  &        7.855 $\pm$  0.228  &        2.578 $\pm$  0.107  &           --         \\
      6 Hebe                &  {\bf 24.248 $\pm$  0.865} &  {\bf 17.382 $\pm$  0.553} &                --          &           --         \\
      6 Hebe (2)            &  {\bf 25.937 $\pm$  0.934} &  {\bf 20.272 $\pm$  0.658} &   {\bf 6.046 $\pm$  0.156} &           --         \\
      511 Davida            &  {\bf 18.076 $\pm$  0.621} &  {\bf 14.896 $\pm$  0.465} &   {\bf 4.966 $\pm$  0.144} &           --         \\
      511 Davida (2)        &  {\bf 17.690 $\pm$  0.606} &  {\bf 13.687 $\pm$  0.423} &                --          &           --         \\
      7 Iris                &       67.823 $\pm$  2.803  &       57.474 $\pm$  2.166  &       32.242 $\pm$  0.727  &  15.984 $\pm$  0.675 \\
      2 Pallas              &  {\bf 62.331 $\pm$  2.544} &  {\bf 46.987 $\pm$  1.719} &  {\bf 17.907 $\pm$  0.414} &           --         \\
      1 Ceres               & {\bf 258.977 $\pm$ 12.991} & {\bf 219.497 $\pm$ 10.082} &  {\bf 72.467 $\pm$  1.619} &  {\bf 55.259 $\pm$  2.286} \\
      93 Minerva            &   {\bf 6.746 $\pm$  0.208} &   {\bf 4.749 $\pm$  0.132} &   {\bf 1.750 $\pm$  0.065} &           --         \\
      65 Cybele             &  {\bf 16.579 $\pm$  0.564} &  {\bf 12.446 $\pm$  0.380} &   {\bf 4.897 $\pm$  0.122} &           --         \\
      4 Vesta               & {\bf 206.512 $\pm$ 10.038} & {\bf 155.047 $\pm$  6.779} &  {\bf 47.163 $\pm$  1.064} &  {\bf 39.749 $\pm$  1.652} \\
      4 Vesta (2)           &      156.430 $\pm$  7.310  &      117.833 $\pm$  4.951  &       37.654 $\pm$  0.887  &           --         \\
      52 Europa             &  {\bf 24.736 $\pm$  0.887} &  {\bf 18.149 $\pm$  0.582} &   {\bf 6.093 $\pm$  0.168} &           --         \\
      52 Europa (2)         &       22.525 $\pm$  0.797  &       17.675 $\pm$  0.565  &        5.879 $\pm$  0.260  &           --         \\
      Neptune               & {\bf 314.872 $\pm$ 16.229} & {\bf 384.606 $\pm$ 19.106} & {\bf 267.687 $\pm$  5.937} & {\bf 238.428 $\pm$  9.814} \\
      Neptune (2)           &      315.032 $\pm$ 16.237  &      366.421 $\pm$ 18.082  &      233.951 $\pm$  5.221  & 199.637 $\pm$  8.253 \\
      47 Aglaja             &   {\bf 6.289 $\pm$  0.193} &                --          &   {\bf 1.600 $\pm$  0.069} &           --         \\
      511 Davida (3)        &  {\bf 20.701 $\pm$  0.723} &  {\bf 17.935 $\pm$  0.572} &                --          &   {\bf 8.568 $\pm$  0.420} \\ \hline

      IRAS 08201$+$2801     &   1.054 $\pm$  0.035 &   1.147 $\pm$  0.032 &           --         &           --         \\
      IRAS 08201$+$2801 (2) &   1.074 $\pm$  0.033 &   1.128 $\pm$  0.031 &           --         &           --         \\
      IRAS 08591$+$5248     &   0.630 $\pm$  0.026 &   1.058 $\pm$  0.030 &           --         &           --         \\
      IRAS 08572$+$3915     &   6.039 $\pm$  0.184 &   5.494 $\pm$  0.154 &    2.085 $\pm$ 0.090 &           --         \\
      IRAS 08474$+$1813     &   1.241 $\pm$  0.037 &   1.495 $\pm$  0.039 &    0.572 $\pm$ 0.062 &           --         \\
      IRAS 08474$+$1813 (2) &   1.197 $\pm$  0.036 &   1.461 $\pm$  0.039 &    1.124 $\pm$ 0.073 &           --         \\
      Arp 220               & 117.838 $\pm$  5.280 & 137.905 $\pm$  5.920 &   86.655 $\pm$ 1.935 &  73.970 $\pm$  3.060 \\
      Mrk 231               &  30.019 $\pm$  1.101 &  28.917 $\pm$  0.983 &   20.411 $\pm$ 0.461 &  17.709 $\pm$  0.738 \\
      IRAS 20100$-$4156     &   4.664 $\pm$  0.139 &   4.896 $\pm$  0.136 &           --         &           --         \\
      UGC 05101             &  10.570 $\pm$  0.340 &  15.788 $\pm$  0.494 &   15.529 $\pm$ 0.358 &  13.973 $\pm$  0.584 \\
      IRAS 00188$-$0856     &   2.616 $\pm$  0.076 &   2.599 $\pm$  0.070 &    2.027 $\pm$ 0.089 &           --         \\
      IRAS 15250$+$3609     &   5.958 $\pm$  0.180 &   4.927 $\pm$  0.136 &           --         &           --         \\
      IRAS 03158$+$4227     &   3.701 $\pm$  0.109 &   3.961 $\pm$  0.109 &           --         &           --         \\ \hline \\
      \multicolumn{5}{@{}l@{}}{\hbox to 0pt{\parbox{180mm}{\footnotesize
            Notes. 
            \par\noindent
            The data designated with the bold-faced were used for the derivation of the calibration factors. 
          }\hss}}
    \end{tabular}
  }
\end{center}
\end{table}

\begin{table}
  \begin{center}
    \caption{Summary of the calibration factors for point-source photometry.}
    \label{tab:abscal}
    \begin{tabular}{lcrr}
      \hline
      Band name     & Calibration factor\footnotemark[$*$]    & Calibration accuracy [\%] & Flux range\footnotemark[$\dagger$] [Jy] \\ \hline
      {\it N60}     & $0.698 \times (\mbox{\rm{total flux}})^{-0.0659} $ & 13.7  & 0.1--300  \\
      {\it WIDE-S}  & $0.700 \times (\mbox{\rm{total flux}})^{-0.0757} $ & 12.7  & 0.1--400  \\
      {\it WIDE-L}  & $0.560 $                                          &  9.97  & 0.5--300  \\
      {\it N160}    & $0.277 $                                          & 50.5  &  10--250  \\ \hline \\
      \multicolumn{4}{@{}l@{}}{\hbox to 0pt{\parbox{180mm}{\footnotesize
            Note. 
            \par\noindent
            \footnotemark[$*$] Total flux is a sum of the observed flux, background sky flux, and the detector dark current. 
            \par\noindent
            \footnotemark[$\dagger$] Confirmed flux range. \\
            \quad In the case of the fainter sources, the extrapolation should be possible, \\
            \quad because the total fluxes are dominated by the detector dark current for the {\it N60} and {\it WIDE-S} bands, \\
            \quad and the offset light signal for the {\it WIDE-L} and {\it N160} bands, respectively. \\
            \quad The brighter end is almost comparable to the saturation limit.     
          }\hss}}
    \end{tabular}
  \end{center}
\end{table}

\begin{table}
  \begin{center}
    \caption{Color correction factors.}
    \label{tab:color}
    {\scriptsize
    \begin{tabular}{lcccc}
      \hline
      Intrinsic     & {\it N60} & {\it WIDE-S} & {\it WIDE-L} & {\it N160} \\
      spectrum      & (65 $\rm{\mu m}$) & (90 $\rm{\mu m}$) & (140 $\rm{\mu m}$) & (160 $\rm{\mu m}$) \\ \hline
      Black-body\footnotemark[$*$] ($\beta=0$) & & & & \\
      - $T=10$      & 4.434  &  1.840  &  1.549  &  1.097  \\
      - $T=30$      & 1.050  &  0.892  &  0.957  &  0.986  \\
      - $T=50$      & 0.976  &  0.979  &  0.937  &  0.986  \\
      - $T=70$      & 0.978  &  1.066  &  0.935  &  0.988  \\
      - $T=100$     & 0.992  &  1.154  &  0.935  &  0.989  \\
      - $T=300$     & 1.029  &  1.320  &  0.936  &  0.992  \\
      - $T=1000$    & 1.044  &  1.381  &  0.937  &  0.993  \\
      - $T=3000$    & 1.048  &  1.398  &  0.937  &  0.993  \\
      - $T=10000$   & 1.049  &  1.404  &  0.937  &  0.993  \\ \hline
      Gray-body\footnotemark[$*$] ($\beta=-1$) & & & & \\
      - $T=10$      & 5.248  &  2.093  &  1.770  &  1.143  \\
      - $T=30$      & 1.107  &  0.902  &  0.999  &  0.994  \\
      - $T=50$      & 0.997  &  0.935  &  0.962  &  0.989  \\
      - $T=70$      & 0.983  &  0.986  &  0.953  &  0.989  \\
      - $T=100$     & 0.985  &  1.041  &  0.949  &  0.989  \\
      - $T=300$     & 1.005  &  1.148  &  0.945  &  0.990  \\
      - $T=1000$    & 1.013  &  1.187  &  0.944  &  0.990  \\
      - $T=3000$    & 1.016  &  1.198  &  0.943  &  0.990  \\
      - $T=10000$   & 1.016  &  1.202  &  0.943  &  0.990  \\ \hline
      Gray-body\footnotemark[$*$] ($\beta=-2$) & & & & \\
      - $T=10$      & 6.281  &  2.396  &  2.048  &  1.198  \\
      - $T=30$      & 1.178  &  0.930  &  1.060  &  1.008  \\
      - $T=50$      & 1.030  &  0.918  &  1.003  &  0.998  \\
      - $T=70$      & 1.001  &  0.944  &  0.987  &  0.995  \\
      - $T=100$     & 0.992  &  0.976  &  0.978  &  0.994  \\
      - $T=300$     & 0.995  &  1.041  &  0.968  &  0.993  \\
      - $T=1000$    & 0.999  &  1.065  &  0.965  &  0.992  \\
      - $T=3000$    & 1.000  &  1.072  &  0.964  &  0.992  \\
      - $T=10000$   & 1.001  &  1.075  &  0.964  &  0.992  \\ \hline
      Power-law\footnotemark[$\dagger$] & & & & \\
      - $\alpha=-3$ & 1.040  &  0.954  &  1.129  &  1.033  \\
      - $\alpha=-2$ & 1.013  &  0.962  &  1.054  &  1.013  \\
      - $\alpha=-1$ & 1.000  &  1.000  &  1.000  &  1.000  \\
      - $\alpha=0$  & 1.001  &  1.076  &  0.964  &  0.992  \\
      - $\alpha=1$  & 1.017  &  1.203  &  0.943  &  0.990  \\
      - $\alpha=2$  & 1.049  &  1.407  &  0.937  &  0.993  \\
      - $\alpha=3$  & 1.101  &  1.724  &  0.945  &  1.001  \\ \hline \\
      \multicolumn{5}{@{}l@{}}{\hbox to 0pt{\parbox{180mm}{\footnotesize
            Note. 
            \par\noindent
            \footnotemark[$*$] Black-body and Gray-body spectrum : $F(\nu) \propto B_{\nu}(T) \cdot \nu^{\beta}$.
            \par\noindent
            \footnotemark[$\dagger$] Power-law spectrum : $F(\nu) \propto \nu^{\alpha}$.
          }\hss}}
    \end{tabular}
  }
  \end{center}
\end{table}

\end{document}